\documentclass{IEEEtran}
\usepackage{cite}

\ifCLASSINFOpdf
  \usepackage[pdftex]{graphicx}
  \graphicspath{{./}}
  \DeclareGraphicsExtensions{.pdf,.jpeg,.png}
\else
   \usepackage[dvips]{graphicx}
   \graphicspath{{./figure/}}
   \DeclareGraphicsExtensions{.eps}
\fi
\usepackage[cmex10]{amsmath}

\newtheorem{definition}{Definition}
\newtheorem{lemma}{Lemma}
\newtheorem{theorem}{Theorem}
\newtheorem{proposition}{Proposition}
\newtheorem{corollary}{Corollary}

\usepackage{algorithm}
\usepackage{amsfonts}
\usepackage{amssymb}
\usepackage{color}
\usepackage{footnote}
\makesavenoteenv{table}
\usepackage[T1]{fontenc}
\usepackage[inline]{enumitem}
\usepackage{epstopdf}

\usepackage[font=footnotesize]{subfig}
\usepackage[hyphens]{url}

\begin{document}
%
\title{Cascading DoS Attacks on IEEE 802.11 Networks}

\author{Liangxiao Xin, David Starobinski, and Guevara Noubir
\thanks{L. Xin, and D. Starobinski are with the Division of Systems Engineering, Boston University, Boston, MA 02215 USA (e-mail: xlx@bu.edu;
staro@bu.edu).}
\thanks{G. Noubir is with the College of Computer and Information Science, Northeastern University, Boston, MA 02115 USA (e-mail:
noubir@ccs.neu.edu).}
}


%


\maketitle

\begin{abstract}
We unveil the existence of a vulnerability in Wi-Fi (802.11) networks, which allows an adversary to remotely launch a Denial-of-Service (DoS)
attack that propagates  both in time and space. This vulnerability stems from a coupling effect induced by hidden nodes. Cascading DoS attacks can
congest an entire network and do not require the adversary to violate any protocol. We demonstrate the feasibility of such attacks through
experiments with real Wi-Fi cards, extensive ns-3 simulations, and theoretical analysis. The simulations show that the attack is effective both in
networks operating under fixed and varying bit rates, as well as ad hoc and infrastructure modes. To gain insight into the root-causes of the
attack, we model the network as a dynamical system and analyze its limiting behavior and stability. The model predicts that a phase transition
(and hence a cascading attack) is possible when the retry limit parameter of Wi-Fi is greater or equal to 7, and characterizes the phase
transition region in terms of the system parameters.
\end{abstract}


%
\IEEEpeerreviewmaketitle

\section{Introduction}
\label{Introduction}
Wi-Fi (IEEE 802.11) is a
technology widely used to access the Internet. Wi-Fi connectivity is provided by
a variety of organizations operating over a shared RF spectrum. These
include schools, libraries, companies,  towns and governments,  as well
as ISP hotspots and residential wireless routers.
Wi-Fi traffic is also rapidly rising due to increased
offloading by cellular operators~\cite{lee2010mobile}.
The importance of Wi-Fi networks and the need to strengthen their resilience to intentional and
non-intentional interference have been recognized by companies, such as
Cisco~\cite{Cisco}.

Wi-Fi networks rely on simple, distributed mechanisms to arbitrate access to the shared spectrum and optimize performance. Such mechanisms include
carrier sensing multiple access (CSMA), exponential back-offs, and bit rate adaptation.  The behavior of these mechanisms in isolated single-hop
networks has been extensively studied and is generally well-understood (see, e.g.,~\cite{bianchi}). However, due to interference coupling, these
mechanisms result in complex interactions in multi-hop settings. As a consequence, different networks do not always evolve independently, even if
they are located far away.

\begin{figure}[!t]
\centering
\includegraphics[width=2.5in]{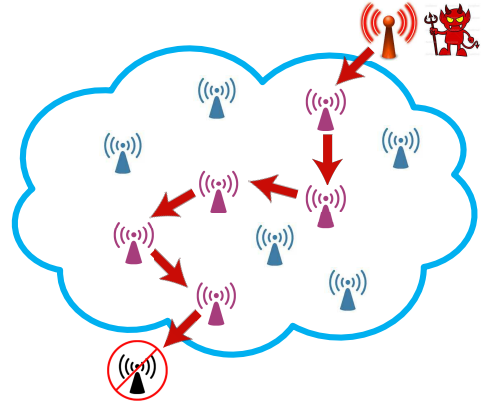}
\caption{Illustration of a cascading denial of service attack. Transmissions by an attacker impact nodes located far away, due to interference
coupling caused by hidden nodes.}
\label{general_case}
\end{figure}

Figure~\ref{general_case} serves to illustrate this phenomenon at a high level. Suppose that an attacker increases the rate at which it generates
packets, and transmits these packets in accordance with the IEEE 802.11 protocol.
These transmissions may cause packet collisions at nodes concurrently receiving packets from other sources. Due to the infamous hidden node problem,
which is hard to avoid in wireless networks, transmitters may be unable to hear transmission by other nodes, even when using CSMA, and hence keep
retransmitting packets until they reach the so-called retry limit of the back-off procedure. These retransmissions affect other neighbours and may
propagate.

While an optional mechanism, called RTS/CTS, has been designed to combat the hidden node problem,
it increases overhead and latency especially
at high bit rates. Since the cost of the RTS/CTS exchange usually does not
justify its benefits, it is commonly
disabled~\cite{forouzan2004data, gast2005802}. Indeed, most manufacturers of Wi-Fi cards disable RTS/CTS by default and discourage changing this
setting as explicitly stated in~\cite{netgear,tp-link,linksys,d-link}.  Therefore, most Wi-Fi systems today operate without RTS/CTS.

The coupling phenomenon induced by interferences
creates multi-hop dependencies, which an adversary can take advantage of to launch a widespread network attack from a single location. We refer to
such an attack as a \emph{cascading Denial-of-Service (DoS) attack}. Cascading DoS attacks are especially dangerous because they affect the entire
network and do not require the adversary to violate any protocol (i.e., the attacks are protocol-compliant).

The contributions of this paper are as follows. First, we unveil the existence of a vulnerability in the IEEE 802.11 standard,
 which allows an attacker to launch protocol-compliant cascading DoS attacks. In contrast to
 existing jamming attacks, the attacker does not need to be in the vicinity of the victims.

Second, we provide a concrete attack that exploits this vulnerability in certain network scenarios.
We demonstrate the attack through experiments on a testbed composed of nodes equipped with real Wi-Fi cards, and through extensive ns-3
simulations.

Third, we show the existence of a \emph{phase transition}. When the packet generation rate of the attacker is
lower than the phase transition point, it has vanishing effect on the rest of the
network. However, once the packet generation rate exceeds the phase transition point, the network becomes entirely congested.
Thus, under a phase transition,  the utilization of a remote node experiences no change until it is suddenly forced to
congestion~\cite{saligrama2006macroscopic}. 

Finally, we introduce a new analytical model that sheds light into the phase transition observed in the simulations and experiments. We apply fixed point theorems to this model.  The analysis predicts for which values of the retry limit a phase transition (and hence a cascading attack)
can occur, and explicitly characterizes the phase transition region in terms of the system parameters.  In particular, we show that a phase
transition can occur for the default value of the retry limit in Wi-Fi, which is~7. We carry out a stability analysis and demonstrate that in the
phase transition region the system must have multiple fixed points, one of which being unstable.


%
%
%

The rest of the paper is organized as follows. In Section~\ref{Related_Work}, we discuss related work.
In Section~\ref{Background}, we provide brief background on Wi-Fi, hidden nodes, and Minstrel, and introduce our network model.
We present and discuss experimental and simulation
results in Section~\ref{Simulations and Experiment}. In
Section~\ref{Analysis}, we present an analytical model that explains the behaviour of the network and the impact of various parameters, and compare
the analytical and simulation results. In Section~\ref{Conclusion}, we conclude the paper and discuss possible mitigation methods.

An earlier and shorter version of this paper appeared in the proceedings of the IEEE Conference on Communications and Network Security (CNS 2016)~\cite{cns2016}. This journal version significantly expands the theoretical analysis, including detailed proofs of all the lemmas and theorems, and
new results on stability analysis and heterogeneous traffic load, all of which can be found in Section~\ref{Analysis}. Moreover, new simulation results for infrastructure networks, networks supporting RTS/CTS, ring networks, 
networks based on a realistic indoor building model, 
and networks with heterogeneous traffic load are presented in Sections~\ref{Simulations} and~\ref{Mitigation}. 

\section{Related Work}
\label{Related_Work}


In general, the main goal of a DoS attack is to make communication impossible
for legitimate users. Within the context of wireless networks, a simple and popular means to
launch a DoS attack is to jam the network with high power transmissions of random bits, hence creating
interferences and congestion. Jamming at the physical layer, together with \emph{anti-jamming} countermeasures, have been extensively studied
(cf.~\cite{poisel2011modern} for a monograph on this subject).

More recently, several works have developed and demonstrated \textit{smart jamming} attacks. These attacks exploit protocol
vulnerabilities across various layers in the stack to achieve high jamming gain and energy
efficiency, and a low probability of detection~\cite{pelechrinis2011denial}.
For instance, \cite{lin2005link} shows that the energy consumption of a smart
jamming attack can be four orders of magnitude lower than continuous
jamming. The works in~\cite{noubir2011robustness,
  orakcal2014jamming} show that several Wi-Fi bit rate adaptation algorithms, such as
SampleRate, ONOE, AMRR, and RARF, are vulnerable
to smart jamming. However, both conventional and smart jamming attacks are usually non-protocol
compliant. Moreover, they require physical proximity. These limitations can be
used to identify and locate the jammer.

In contrast, in this work we show how a protocol-compliant DoS attack
can be remotely launched by exploiting coupling due to hidden nodes in
Wi-Fi. Rate adaptation algorithms further amplify this attack due to
their inability to distinguish between collisions, interferences, and
poor channels. One potential mitigation is to design a rate
adaptation algorithm whose behaviour is based on the observed interference patterns~\cite{chen2007rate, rayanchu2008diagnosing}.  However, to the
best of
our knowledge, none of these rate adaptation algorithms are used in
practice. Our work is based on Minstrel~\cite{Minstrel}, which is the
most recent, popular, and robust rate adaptation algorithm for Linux systems.

The attacks that we are investigating bear similarity to cascading
  failures in power transmission systems~\cite{kinney2005modeling, soltan2014cascading}. When one of the nodes
in the system fails, it shifts its load to adjacent nodes. These nodes in
turn can be overloaded and shift their load further. This phenomenon
has also been studied in wireless networks. For
instance,~\cite{haenggi2009stochastic, kong2009wireless} model
wireless networks as a random geometric graph topology generated by a
Poisson point process. They use percolation theory to show that the
redistribution of load induces a phase transition in the
network connectivity. However, the cascading phenomenon that we investigate
in this paper is different from cascading failure studied in those works. In our work, the exogenous generation of
traffic at each node is independent. That is, a node will
not shift its load to other nodes. The amount of traffic measured on the channel increases
due to  packet retransmissions caused by packet collisions, rather
than due to traffic redistribution.


The work in~\cite{aziz2009ez,aziz2011understanding} show that interference coupling can affect the stability of multi-hop networks. In the case of a greedy
source, a three-hop network is stable while a four-hop network becomes unstable.
In contrast, in our work, the path of each packet consists of a single-hop. Thus, network instability is not due to multi-hop communication in our case.

The work in~\cite{ray2005performance,saligrama2006macroscopic} show that local coupling due to interferences
can have global effects on wireless networks. Thus, \cite{ray2005performance} proposes
a queuing-theoretic analysis and approximation to predict the probability of a packet
collision in a multi-hop network with hidden nodes. It shows that the sequence of the
packet collision probabilities in a linear network converges to a
fixed point. The work in~\cite{saligrama2006macroscopic} evaluates the impact of rate adaption and finds out that traffic increase
at a single node can congest an entire network, and points out the existence of a phase transition.

Our paper differs in several aspects. First, it considers an adversarial context, and shows how interference-induced coupling can be exploited to
cause denial of service. Second, to our knowledge, it is the first work to demonstrate the existence of such coupling on real commodity hardware.
Third, our simulations are based on a high-fidelity wireless simulator (ns-3), capable of capturing the effects of rate adaptation algorithms and
accurately modeling  infrastructure networks. Finally, our analytical model is original and captures the impact of the retry limit and traffic
parameters. A key result is that a cascading attack can be launched for the default value of the retry limit in Wi-Fi, a result validated by the
experiments and simulations.

\section{Background and Model}
\label{Background}
We first review key aspects of IEEE 802.11 and then describe the network model under consideration.

\subsection{Wi-Fi Summary}
\label{Wi-Fi Summary}
Wi-Fi is a wireless local area network (WLAN) technology, which mainly
runs on 2.4~GHz ISM bands and 5~GHz bands~\cite{gast2005802}. The IEEE 802.11 standard is a
series of specifications, such as the media access control (MAC) and
physical layer (PHY) interfaces. The first 802.11 standard
that gained widespread success is 802.11b.
It runs on 2.4~GHz bands and
has up to 11 Mb/s bit rate.
The subsequent standards (e.g., 802.11a, g, n, and
ac) increased the bit rates using higher order modulation along with
coding, OFDM, MIMO, and wider bands.
It is noteworthy
that 802.11b is the only mode that supports communication at 1~Mb/s. Hence, when the
bit rate reduces to 1 Mb/s, Wi-Fi network reverts to the
802.11b mode. Generally, this lower bit rate has higher resistance to
interference during transmission and is able to operate over lower SNR channels.

The IEEE 802.11 standard uses a CSMA/CA mechanism to control access to
the transmission medium and avoid collisions.
After a packet is sent, a
node waits for a short interframe slots (SIFS) period to receive an
ACK. Whenever the channel becomes idle, the node waits for a distributed
interframe space (DIFS $>$ SIFS) period and a random backoff before contending for the channel.
The random backoff consists of a random number of backoff slots, which depends on the so-called contention window.
Specifically, at the $r \geq 1$ retransmission attempt (retry count), the contention window $CW_r$ is given by
\begin{eqnarray}
CW_r = \left\lbrace
	\begin{array}{ll}
		2^{r-1} (CW_1 + 1) - 1   & CW_r < CW_{max}, \\
		CW_{max}  & \text{otherwise}.
	\end{array}
\right.\label{eq:back-off}
\end{eqnarray}
The number of backoff slots is chosen uniformly at random in the interval $[0, CW_r]$.
For IEEE 802.11b, the initial contention window size is $CW_1 = 31$ , the maximum contention window size is $CW_{max} = 1023$, and the duration of a
backoff slot is $20~\mu s$.
Note that the case $r=1$ corresponds to the initial packet transmission attempt. 
%

\subsection{Hidden Node Problem}
\label{Hidden Node Problem}
\begin{figure}[!t]
\centering
\includegraphics[width=2.5in]{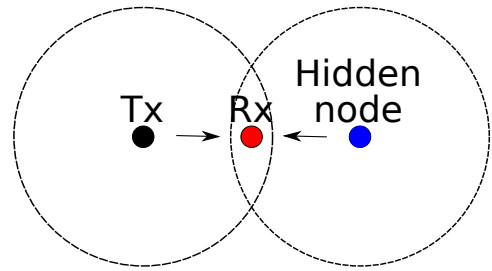}
\caption{Classical hidden node problem. The transmitter and the hidden
  node cannot sense each other. The collision happens when they
  transmit simultaneously.}
\label{classic_hidden_node}
\end{figure}

A typical instance of the hidden node problem is illustrated in
Figure~\ref{classic_hidden_node}. The figure shows three nodes: a transmitter,
a receiver and a hidden node. The dashed circle represents the
transmission range of the node. Since the transmitter and the hidden
node cannot sense each other, a collision happens when both of them
transmit packets at the same time.

A packet collision triggers a
retransmission. In IEEE 802.11, there is an upper limit on the number
of retransmissions that a packet can incur, called \textit{retry limit} and
denoted by $R$ (the default value is $R=7$). If the retry count $r$ of a packet exceeds the retry
limit, the packet is dropped, the retry count is reset to $r=1$, and a new packet transmission can start.  The channel utilization of a node increases with the
probability of a packet collision. In the worst case,
the utilization can be~$R$ times larger than in the absence of packet collisions. Therefore, the access channel of
a node can easily be saturated if it is forced to retransmit packets.

The hidden node problem can in principle be avoided by enabling the RTS/CTS exchange,
which is implemented in Wi-Fi networks. However, the RTS/CTS exchange has not only high overhead, but also does not always fully prevent packet
collisions~\cite{ray2005evaluation}
and may lead to deadlocks in multi-hop configurations~\cite{ray2007false}.
Generally, it is either turned off
\cite{bellardo2003802} or only used for packets whose length exceeds the so-called RTS
threshold. Most manufacturers of Wi-Fi cards, including Netgear~\cite{netgear}, TP-LINK~\cite{tp-link}, Linksys~\cite{linksys} and
D-Link~\cite{d-link}, disable RTS/CTS altogether by setting the RTS
threshold to a sufficiently high default value (e.g., 2346~bytes, which corresponds to the maximum length of an IEEE~802.11 frame).
They furthermore recommend to not change the default setting.

\subsection{Minstrel Rate Adaptation}
\label{Minstrel Rate Adaptation}
Minstrel is a practical, state-of-the-art rate adaptation algorithm that has been
implemented within the MadWiFi project and Linux mac80211
driver framework~\cite{Minstrel}. It chooses the bit rate of a
transmission based on the throughput measured over past transmissions at different rates. Technically, it
selects a bit rate following a retry chain, as shown in
Table~\ref{Minstrel retry chain}.

In Minstrel, 90\% of the packets are
transmitted at a ``normal rate'' (fourth column in Table~\ref{Minstrel retry chain}). The remaining 10\% are
transmitted at a ``lookaround rate'' (second and third columns in Table~\ref{Minstrel retry chain}). Each packet is transmitted at a
rate following a retry chain (rows in Table~\ref{Minstrel retry
  chain}). For example, consider a packet being transmitted at
``lookaround rate''. If a random rate is lower than the rate with ``best
throughput'', the packet is first transmitted at the
``best throughput'' rate, then at the ``random rate'', then at the ``best probability'' rate,
and finally at the ``lowest baserate''. The
packet is dropped if the transmission fails at the ``lowest
baserate''. The retry chain table is updated 10 times every second
based on performance statistics.

Therefore, a large amount of packet loss does not
necessarily cause Minstrel to switch to a low bit rate. Another advantage of
Minstrel is that it probes the throughput of different bit rates
randomly. This makes the rate adaptation more robust in complicated
environment and against some adversaries.

\begin{table}
\small
\centering
\caption{Minstrel Retry Chain \cite{Minstrel}}
\begin{tabular}{|c|c|c|c|} \hline \label{Minstrel retry chain}
Try & \multicolumn{2}{|c|}{Lookaround rate} & Normal rate \\
\cline{2-3}
    & random $<$ best\footnote{The random rate is lower than the best throughput rate.} & random $>$ best    & \\
\hline
1   & Best throughput  & Random rate      & Best throughput \\
\hline
2   & Random rate      & Best throughput  & 2nd best throughput \\
\hline
3   & Best probability\footnote{This rate has the highest probability of resulting in a successful transmission.} & Best probability & Best
probability \\
\hline
4   & Lowest baserate & Lowest baserate & Lowest baserate \\
\hline
\end{tabular}
\end{table}

\subsection{Network Model}
\label{Network Model}
\begin{figure}[!t]
\centering
\includegraphics[width=3.2in]{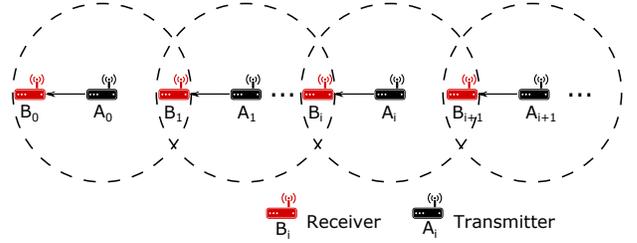}
\caption{Topology of the network. Node $A_i$ transmits packets to node $B_i$. Node $A_{i}$ is a hidden node with respect to $A_{i+1}$.}
\label{linear_topology}
\end{figure}

The network model considered in this paper is shown in Figure~\ref{linear_topology}. This configuration could arise over different time and space
in
more complex network topologies. We consider
 $N+1$ pairs of nodes.  Each node $A_i$ ($i=0, 1, 2, \ldots$, $N$)
transmits packets to node $B_i$. The dashed circle represents the
range of transmission. Node $B_{i+1}$ can receive packets from both node
$A_i$ and node $A_{i+1}$. However, node $A_i$ and node $A_{i+1}$
cannot hear each other. That is, node $A_i$ is a hidden node with
respect to node $A_{i+1}$ (and vice-versa). A packet collision happens at node
$B_{i+1}$ when packet transmissions by node $A_i$ and $A_{i+1}$ overlap.

We assume that all the nodes communicate over the same channel. Note that there are only three non-overlapping channels in the 2.4GHz band.
Hence, it is common that several nodes use the same channel over time and space in crowded areas.


\subsection{Cascading DoS attack}
Our goal is to investigate how node $A_0$ can trigger a cascading DoS
attack, resulting in a congestion collapse over the entire network. We start
by increasing the packet generation rate at node
$A_0$. Node $A_0$ transmits packets over its channel, in compliance with the IEEE 802.11 standard. The transmissions by
node $A_0$ cause packet collisions at
node $B_1$. These collisions require node $A_1$ to retransmit packets. The increased  amount of packet transmissions and retransmissions by node
$A_1$ impact node $A_2$ and so forth. If this effect keeps propagating and amplifying, then the result is a network-wide  denial of service, which
we refer to as a cascading Denial of Service (DoS) attack. Because this attack is protocol-compliant, it is difficult to detect or trace back to the
initiator.

We note here that as  a hidden node retransmits its packets, it must back off after each retransmission, which leaves the channel idle for a certain
period of time. However, the duration of the backoff period is generally too short to allow for a successful transmission.
Indeed, a packet transmission is successful only if
\begin{enumerate}
    \item The size of the contention window  of the hidden node is longer than the packet transmission time.
    \item The transmitter starts and ends its transmission entirely during the backoff period of the hidden node.
\end{enumerate}
At 1~Mb/s, the transmission time of an 1500~bytes packet lasts 12~ms. This is longer than
the contention window as long as $CW_{r} < CW_{max} = 1023$. Hence, by Eq.~(\ref{eq:back-off}), a transmission cannot be successful during the
backoff period preceding the $r < 6$ retransmission attempt by a hidden node.

At the $r \geq 6$ retransmission attempt by a hidden node $A_i$, $CW_{r} = CW_{max} = 1023$. Node $A_i$  back-offs for $n$ slots, where $n$ is an integer between 0 and 1023 that is picked uniformly at random (i.e., with probability $1/1024$).
Since the length of a backoff slot is 20~$\mu$s, the backoff delay is $0.02n$~ms. Without loss of generality, assume that node~$A_i$ starts backing off at time $t=0$ and ends its backoff at time $t=0.02n$ (all the time units are in milliseconds). Node $A_i$ then starts a packet transmission, which ends at time $t=0.02n+0.12$.

Node $A_{i+1}$ can transmit a packet successfully only if it starts its transmission during the time interval $[0,0.02n-12]$. This requires $n>600$. Assuming that the starting time of the packet transmission by node $A_{i+1}$ is uniformly distributed in the time interval $[0,0.02n+12]$, the probability that a packet is successfully transmitted by node $A_{i+1}$  is \[\sum_{n=600}^{1023} \frac{1}{1024} \cdot \frac{0.02n-12}{0.02n+12} = 0.059.\]
Thus, the likelihood of a successful packet transmission is low, a result validated by the experimental and simulation results of the next section.

\section{Experimental and Simulation Results}
\label{Simulations and Experiment}
In this section, we demonstrate the feasibility of launching cascading DoS attacks both
through experiments and simulations. We first show results on an experimental testbed using real Wi-Fi cards.
We then use ns-3.22  simulations to investigate how this attack can be performed in significantly larger
scale networks, and under different settings (ad~hoc, infrastructure, fixed bit rate, and adaptive bit rate).

\subsection{Experiments}
\label{Experiment}
\begin{figure}[!t]
\centering
\includegraphics[width=2.5in]{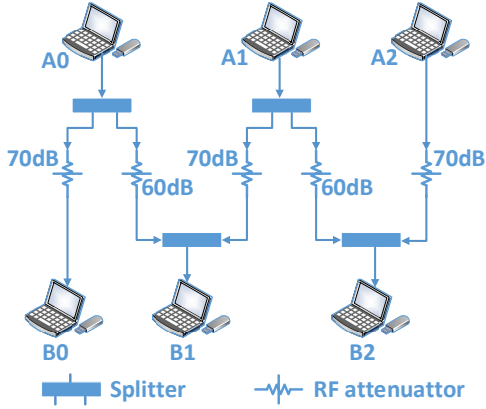}
\caption{Experimental testbed.}
\label{exp_setup}
\end{figure}

 We set up an experimental testbed composed of six nodes. The testbed
 configuration is shown in Figure~\ref{exp_setup}. We establish an
IEEE 802.11n ad hoc network consisting of three pairs of nodes. Each node
consists of a PC and a TP-LINK TL-WN722N Wireless USB Adapter. We use
RF cables and splitters to link the nodes, isolate them from
external traffic, and obtain reproducible results.

We place 70~dB attenuators on links between node $A_i$ and $B_i$ ($i \in 0,1,2$), and 60~dB attenuators
on links between nodes $A_i$ and $B_{i+1}$. The difference in the
signal attenuation of different links ensures that a packet loss occurs if a hidden node transmits. In practice, such a
situation may occur if nodes $A_i$ and $B_{i+1}$ communicate without obstacles, while node $A_i$ and $B_i$ are separated by an office
wall~\cite{stein1998indoor}.
The transmission power of each node is set to 0~dBm.
We use iPerf
\cite{IPERF} to generate UDP data streams and to measure the throughput achieved on
each node. The length of a packet is the default IP packet size of
1500 bytes.

\begin{figure}[!t]
\centering
\includegraphics[width=2.5in]{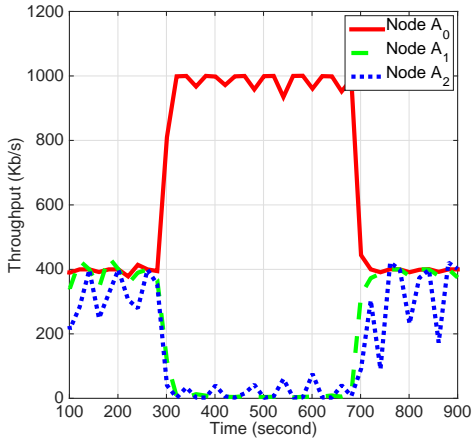}
\caption{Throughput performance measurements in testbed. When node
  $A_0$ starts increasing its packet generation rate, the throughput of nodes $A_1$
  and $A_2$ vanishes.}
\label{implementation_result}
\end{figure}

Figure~\ref{implementation_result} demonstrates the cascading DoS
attack on the experimental testbed. At first,  the packet generation
rates of  nodes $A_0, A_1$ and $A_2$ are set to 400 Kb/s. We observe that the throughput of all the nodes
remains in the vicinity of 400 Kb/s during the first 300 seconds. After 300~seconds,
$A_0$ starts transmitting packets at 1 Mb/s. As a result, the throughput of nodes $A_1$ and $A_2$ suddenly
vanishes.  Once node $A_0$ resumes transmitting at 400 Kb/s, the
throughput of node $A_1$ and node $A_2$ recovers.

%

\subsection{Simulations}
\label{Simulations}
In the previous section, we demonstrated the
feasibility of launching a cascading DoS attack on an experimental
testbed. This testbed relies on commercial cards that are
black boxes for all purposes.  For instance, the driver of the Wi-Fi card
and the rate adaptation algorithm are closed-source. There are also
substantial usage restrictions, such as parameter settings.

In order to gain a better insight into the
attack in large-scale networks, we resort to ns-3 simulations, a state-of-the-art simulator which includes high-fidelity wireless libraries. We show
the occurrence of cascading DoS attacks
\begin{enumerate}
  \item In ad hoc networks with fixed bit rate;
  \item In ad hoc networks under Minstrel rate adaptation;
  \item In infrastructure networks;
  \item In ring topology networks;
  \item In an indoor scenario;
\end{enumerate}
and the countering of cascading DoS attacks
\begin{enumerate}
  \setcounter{enumi}{5}
  \item In networks with RTS/CTS enabled.
\end{enumerate}

\begin{figure}[!t]
\centering
\subfloat[As the traffic load at node $A_0$ increases, the
  utilization of remote nodes (e.g., $A_{20}$ and $A_{40}$) exhibits a
  phase transition.]
           {\includegraphics[width=3in]{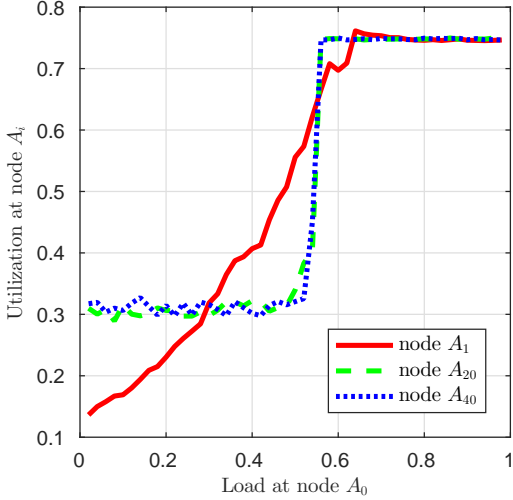}
\label{one_node_phase_transaction}}
\vfil \subfloat[Utilization of nodes $A_i$ ($i \geq 1$) for
  different traffic loads at node $A_0$. The utilization converges as $i$ gets large. When the load at node $A_0$ changes from $0.4$ to
  $0.6$, the sequence of utilization converge to different limits, illustrating the phase transition.]
      {\includegraphics[width=3in]{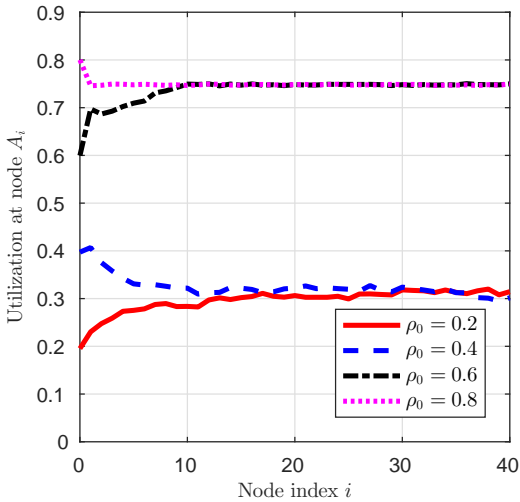}
\label{one_node_load_phase_transaction}}
\caption{Occurrence of cascading DoS attacks in ad hoc networks with fixed bit rate.}
\label{The occurrence of cascading DoS attack in adhoc networks with fixed bitrate}
\end{figure}

\subsubsection{Fixed bit rate}
\label{EXP:Fixed bitrate}
We first describe the occurrence of a cascading DoS attack in an
ad~hoc network with fixed bit rate. We consider a linear topology
consisting of 41 pairs of nodes (i.e. a sequence of 41 hidden nodes), as shown in Figure~\ref{linear_topology}.
Each packet is transmitted over a single-hop path (similar to Wi-Fi Direct).
We fix the bit rate to 1 Mb/s and the retry limit to $R=7$.

We set up a Wi-Fi network using the standard IEEE 802.11 library in ns-3.  At each node $A_i$, $i \geq 1$,
 the generation rate of UDP packets is $\lambda_i=8.125$ pkts/s. The
generation rate of UDP packets at node $A_0$, $\lambda_0$, varies from $1.25$ to $61.25$
pkts/s. Packets at each node are generated according to a Poisson process, hence different nodes start transmitting at different times.
The size
of each packet is 2000 bytes. Each node has the same transmission power
(40~mW).
We set the propagation loss between node $A_i$ and $B_i$ to
80~dB and the propagation loss between node $A_i$ and $B_{i+1}$
to 70~dB.
We run each simulation five times for 1,000 seconds, and average out the results.

The \emph{(exogenous) load} at each node $A_i$ is denoted $\rho_i=\lambda_i T$, where $T$ represents the duration of each packet transmission
attempt (0.016 second in our case).
The \emph{utilization} of a node $A_i$, denoted $u_i$, is defined as the
fraction of time the node is busy transmitting bits on the channel.



Figure~\ref{The occurrence of cascading DoS attack in adhoc networks  with fixed bitrate}(a) depicts the utilization
$u_1$, $u_{20}$, and $u_{40}$ as a function of
$\rho_0$, the load at node $A_0$. The utilization of node~$A_1$, $u_1$, increases smoothly until it reaches its
upper limit. However, the
utilizations of nodes $A_{20}$ and $A_{40}$  remain low until $u_0$ reaches a certain
threshold around $\rho_0 = 0.5$, at which point $u_{20}$ and $u_{40}$ suddenly jump
to a high value.  This sudden jump corresponds to a phase transition, and the critical threshold represents the phase transition point.


Figure~\ref{The occurrence of cascading DoS attack in adhoc networks  with fixed bitrate}(b) illustrates the phase transition in a different way.
The figure depicts the
utilization of each node $A_i$ for $i \geq 1$, as $i$ increases. Again, we observe that different values of
$\rho_0$ lead to two completely distinct behaviour for the sequence of
utilizations $(u_i)_{i=0}^{40}$ (i.e., $u_{40} \simeq 0.3$ when $\rho_0 = 0.2$ and $\rho_0 = 0.4$, while
$u_{40} \simeq 0.75$ when $\rho_0 = 0.6$ and $\rho_0 = 0.8$). 
 Note that the upper limit of the utilization does not reach~1,
due to inter-frame spacing requirements  and (random) backoff delays mandated by IEEE 802.11.

\begin{figure}[!t]
\centering
\subfloat[Throughput]{\includegraphics[width=2.5in]{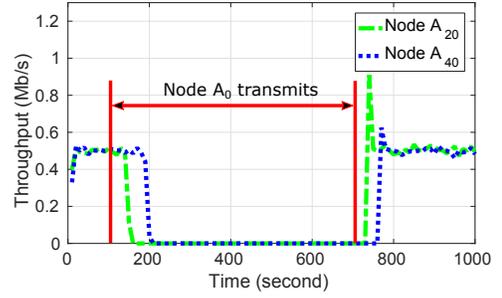}
\label{Minstrel_PHYRateMode_default_onoff_throughput}}
\vfil
\subfloat[Bit rate]{\includegraphics[width=2.5in]{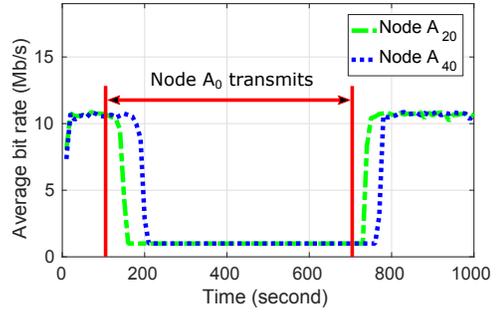}
\label{Minstrel_PHYRateMode_default}}
\caption{Simulation results with Minstrel rate adaptation. When node
  $A_0$ generates packets at 5~Mb/s and transmits, the throughput of nodes $A_{20}$ and $A_{40}$
  vanishes. The average bit rates of nodes $A_{20}$ and $A_{40}$ also
  reduce to 1 Mb/s. This result indicates that nodes $A_{20}$ and
  $A_{40}$ are transmitting packets at the lowest bit rate,
  however with no throughput (all their packets collide).}
\label{Simulation Result with Minstrel Rate Adaptation}
\end{figure}

\subsubsection{Rate Adaptation}
\label{Rate Adaption}
We next consider the same network setting as in the previous section,
but this time we assume that nodes can transmit at different bit rates.
We specifically assume that nodes implement the Minstrel rate adaptation algorithm.
In this case, the attack works by coercing the rate adaptation algorithm to reduce the bit rate
to 1 Mb/s at each node, thus leading to similar results to those shown in Section
\ref{EXP:Fixed bitrate}. In our simulations, the parameter $\textit{EWMA}$ of Minstrel is set to 0.25 \cite{xia2013evaluation}.

We set $\lambda_0 =312.5$ pkts/s and
$\lambda_i = 31.25$ pkts/s ($i \geq 1$) for the packet generation rates. As shown in Figure \ref{Simulation Result with Minstrel Rate Adaptation},
packet transmissions at
node $A_0$ start after $t=100$~s.
%
During the first 100 seconds, the throughput of nodes $A_{20}$ and
$A_{40}$ remain around 0.5 Mb/s, which implies that all the packets are
received.
Once node $A_0$ starts transmitting packets,
the throughput of nodes $A_{20}$ and $A_{40}$ is brought down to close
to zero. We also observe that the bit rates at node $A_{20}$ and $A_{40}$ go down to 1 Mb/s,
due to the repeated packet collisions. Once node
$A_0$ stops transmitting at $t=700$~s, nodes $A_{20}$ and $A_{40}$ recover.

\begin{figure}[!t]
\centering
\subfloat[When
  node $A_0$ generates packets at 5~Mb/s and transmits, the throughput of nodes $A_{20}$ and $A_{40}$
  vanishes.]{\includegraphics[width=2.5in]{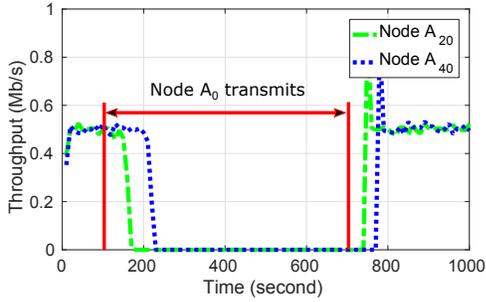}
\label{Simulation Result under AP mode}}
\vfil
\subfloat[When
  node $A_{20}$ generates packets at 5~Mb/s and transmits, the throughput of node $A_{40}$
  vanishes while the throughput of node $A_0$ does not.]{\includegraphics[width=2.5in]{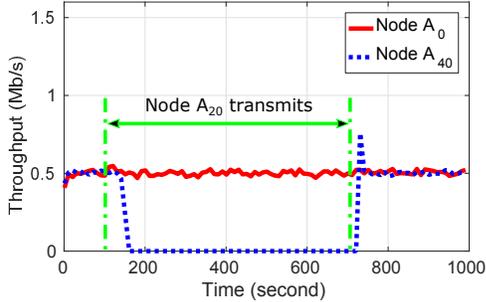}
\label{Simulation Result under AP mode start at middle}}
\caption{Simulation results under AP mode without reassociation. Nodes $A_i$ are stations and
  nodes $B_i$ are access points, for $i \in \{0,1,2,\dots\}$.}
\label{Simulation Result under AP}
\end{figure}

\begin{figure}[!t]
\centering
\includegraphics[width=2.5in]{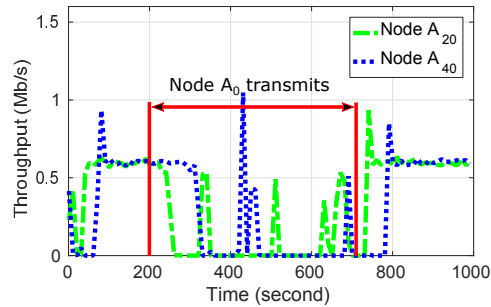}
\caption{Simulation results under AP mode with reassociation. When
  node $A_{0}$ generates packets at 5~Mb/s and transmits, the throughput of node $A_{20}$ and $A_{40}$ significantly decreases.}
\label{Simulation Result under AP mode with reassociation}
\end{figure}

\subsubsection{Infrastructure networks}
We next show that cascading DoS attacks are also feasible in
infrastructure networks. Since the infrastructure mode is more
widely used than ad hoc in practice, the feasibility of the cascading
DoS attack in infrastructure networks increases its severity and
potential impact. We repeat the simulations of Section~\ref{Rate
  Adaption} except that we set nodes $B_i$ as access points, and
nodes $A_i$ as stations. The initial beacon starting time at each AP is a random variable that is uniformly distributed between 0 and $102.4$~ms.

We first investigate the cases where stations do not restart association when beacons are missing.
Toward this end, we set the number of consecutive beacons that must be missed before restarting association, i.e. the attribute {\tt MaxMissBeacons} in ns-3, to a large value.
Otherwise, we use the default settings of ns-3 for the APs \cite{ns3ap} and the stations \cite{ns3sta}. Figure~\ref{Simulation Result under
AP} shows similar results as
in Section~\ref{Rate Adaption}, namely when a cascading DoS attack is launched by node $A_0$, as shown in Figure~\ref{Simulation Result under AP}(a),
the remote nodes $A_{20}$ and $A_{40}$ in the sequence exhibit a phase transition.
If the attacker is node $A_{20}$, the simulation result in Figure~\ref{Simulation Result under AP}(b) shows that the throughput of node
$A_{40}$ vanishes but the throughput of node $A_0$ does not. This result shows that an attack can be launched from any node $A_i$ in the topology
and the following nodes in the sequence (i.e., $A_{i+1}, A_{i+2}, \ldots$) will experience congestion.

We next consider the case where stations restart association when beacons are missing. We set
${\tt MaxMissBeacons} = 10$, which is the default value in ns-3 \cite{ns3sta}. The simulation results are shown in Figure~\ref{Simulation Result under AP mode with reassociation}. When Node $A_0$ starts to transmit packets, we observe a significant throughput degradation at nodes $A_{20}$ and $A_{40}$, but
the throughput does not vanish completely. The reason is that if $A_i$ disassociates from its AP $B_i$ over a certain period
then node $A_{i+1}$ is not affected by interference coupling during that period. This result indicates that reassociations help mitigate cascading DoS attacks, though throughput performance is still significantly impaired.

\subsubsection{Ring topology}
\label{Ring topology}
\begin{figure}[!t]
\centering
\includegraphics[width=2.5in]{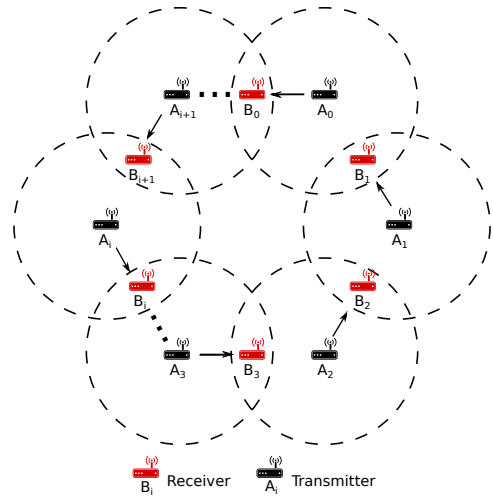}
\caption{Ring topology under cascading DoS attack. The dash circle represents the transmission range of the transmitter.}
\label{ring_topology}
\end{figure}

We investigate cascading DoS attacks in a ring topology with 41 pairs of nodes, as shown in Figure~\ref{ring_topology}. In our previous results for
linear topologies, the effect of an attack disappears once the attacker reduces its packet generation rate. However, the effect of an attack in a
ring topology can last for a long period of time after the attack stops.
Node $A_i$ $(i = 0, 1, \dots)$ generate packets at rate 0.5~Mb/s, following a Poisson process. At time $t=300$ s, node $A_0$ increases its packet generation rate to 11~Mb/s and the
throughput of all the nodes vanishes. Yet, unlike results in linear topologies, the throughput of the nodes does not recover after node $A_0$
reduces its packet generation rate back to 0.5~Mb/s. The cyclic nature of the topology reinforces the attack even after the trigger stops.

\begin{figure}[!t]
\centering
\includegraphics[width=2.5in]{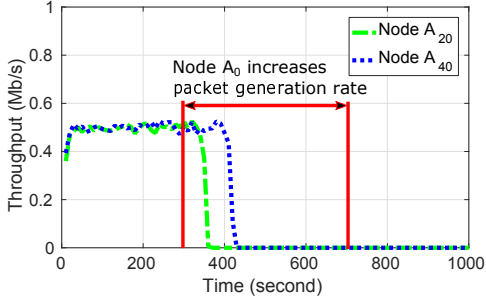}
\caption{Simulation results under a ring topology. When the packet generation rate of node
  $A_0$ increases, the throughput of nodes $A_{20}$ and $A_{40}$
  vanishes. This effect continues even when the packet generation rate of node $A_0$ decreases.}
\label{Simulation Result with Minstrel Rate Adaptation circle}
\end{figure}

This result is illustrated in Figure~\ref{Simulation Result with Minstrel Rate Adaptation circle}.
During the first 100 seconds, all the nodes $A_i$ $(i = 0, 1, \dots)$ generate packets at 0.5~Mb/s. At time $t=300$~s, node $A_0$ increases its packet generation
rate to 11~Mb/s. As a result, the throughput of all nodes vanishes. Yet, unlike results in linear topologies, the throughput of the nodes does not
recover
after node $A_0$ reduces its packet generation rate back to 0.5~Mb/s. The cyclic nature of the topology reinforces the attack even after the trigger
stops.

\subsubsection{Building model} \label{Building model}
In this section, we use the ns-3  {\tt HybridBuildingsPropagationLossModel} library~\cite{ns3hybridbuildingspropagationlossmodel} to demonstrate the feasibility of cascading DoS attacks in an indoor scenario. Models in this library realistically characterize the propagation loss across different spectrum bands (i.e., ranging from 200~MHz to 2.6~GHz), different environments (i.e., urban, suburban, open areas), and different node positions with respect to buildings (i.e., indoor, outdoor and hybrid). The building models take into account the penetration losses of the walls and floors, based on the type of buildings (i.e., residential, office, and commercial).

\begin{figure}[!t]
\centering
\subfloat[Top view.]{\includegraphics[width=2.5in]{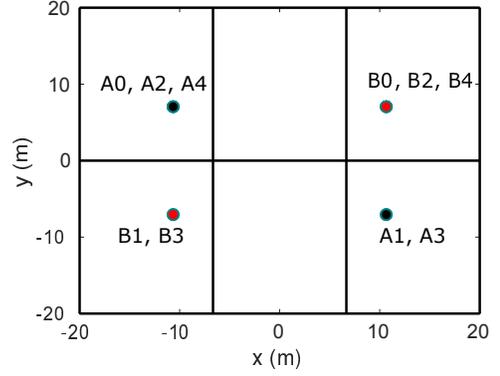}
}
\vfil
\subfloat[Side view.]{\includegraphics[width=2.5in]{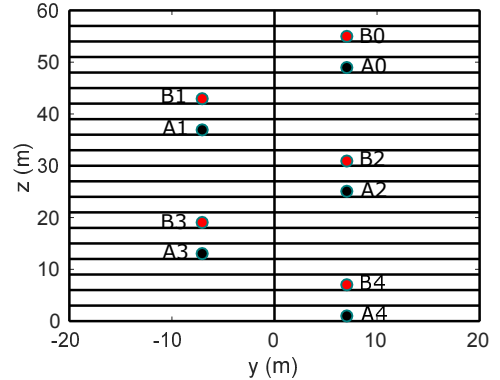}
}
\caption{Office building model. The building has 20 floors ($z$-axis) and 6 rooms in each floor ($x$ and $y$ axes).}
\label{building_model}
\end{figure}

In our simulations, we consider a 20-floor office building with six rooms in each floor, as shown in Figure~\ref{building_model}.  We assume that five pairs of Wi-Fi nodes $(A_i, B_i)$ are active in the building, where node $A_i$ transmits packets to nodes $B_i$ ($i=0,1,2,3,4$). The bit rate is set to 1~Mb/s, the retry limit to $R=7$, and the frequency to 2.4 GHz.  The generation rate of UDP packets at nodes $A_i$, $i \geq 1$,  is $\lambda_i = 8.125$ pkts/s.  Packets are 2000~bytes long.

\begin{figure}[!t]
\centering
\includegraphics[width=2.5in]{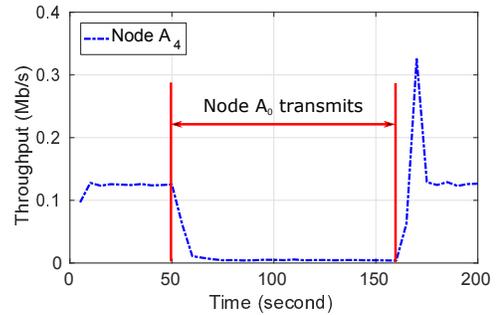}
\caption{Simulation results using ns-3 building model. When node $A_0$ transmits, the throughput of remote node $A_4$ collapses.}
\label{CDoS-1Mbps-adhoc-UDP-building-utilization}
\end{figure}

We turn on and off transmissions at node $A_0$ to observe how it impacts the throughput of other nodes. 
Simulation results are shown in Figure~\ref{CDoS-1Mbps-adhoc-UDP-building-utilization}. When node $A_0$ does not transmit, the throughput of node $A_4$ is 0.13~Mb/s and it incurs no packet loss. However, when node $A_0$ starts transmitting, the throughput of node $A_4$ collapses. The throughput of node $A_4$ recovers only after node $A_0$ stops transmitting.

\subsubsection{RTS/CTS}
\begin{figure}[!t]
\centering
\includegraphics[width=2.5in]{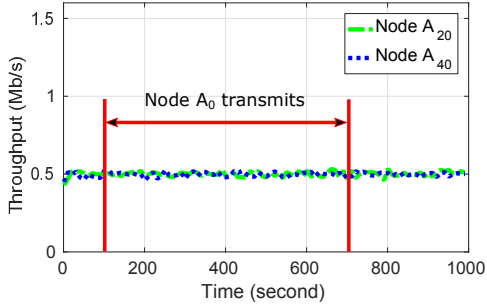}
\caption{Simulation results when enable RTS/CTS. The increase of the packet generation rate of node $A_0$ does not affect the throughput of nodes
$A_{20}$ and $A_{40}$.}
\label{Simulation Result with RTSCTS}
\end{figure}

We next evaluate the impact of enabling RTS/CTS in the topology under consideration. Specifically, we repeat the simulations of Section~\ref{Rate Adaption}, but with RTS/CTS enabled. Figure~\ref{Simulation Result with RTSCTS} shows  that transmissions by node $A_0$, which start after $100$~s, have no effect on the throughput of remote nodes $A_{20}$ and $A_{40}$. This shows that RTS/CTS is an effective solution against cascading DoS attacks in this scenario.

\section{Analysis}
\label{Analysis}
In this section, we develop a stylized, analytical model that provides qualitative insight into the network behavior observed in the simulations and experiments for the
linear topology. Specifically, our goal is to explain why  and under what conditions the phase transition occurs, and shed light into the roles
played by the retry limit~$R$ and the traffic load at the different nodes.

\subsection{Model}
\label{queueing_model}

We consider the linear topology shown in Figure~\ref{linear_topology}. Packet generations at each node $A_i$ form a
Poisson process with rate~$\lambda_i$. The packet size is fixed and the duration of each packet transmission attempt is $T$ (we assume a fixed bit
rate). A transmission by node $A_{i+1}$ is successful only if does not overlap with any transmission by (hidden) node $A_i$.

If a packet collides, it is retransmitted until either it is successfully received or the retry count reaches the limit $R$.  Let $1 \leq \overline{r}_i \leq R$
represent the mean retry count at node $A_i$. Note that the initial packet transmission is included in that count.   Then, the mean service time of a packet at node $A_i$ is $\overline{r}_i T$. To keep the analysis
tractable, timing details of Wi-Fi, such as DIFS, SIFS, and back-off inter-frame spacing are ignored. Therefore the upper limit of the utilization equals 1 in our analysis.

We denote the utilization of node $A_i$ by $0 \leq u_i \leq 1$, where $u_i$ represents the fraction of time node $A_i$ transmits. If $u_i=1$,  node
$A_i$ is congested and transmits continuously. Otherwise, node $A_i$ is uncongested and transmits packets at rate $\overline{r}_i \lambda$.
Therefore, the utilization of node $A_i$ for all $i \geq 0$ is
\begin{equation} \label{definition of utilization}
u_i=\min\{\overline{r}_i \lambda_i T, 1\}.
\end{equation}
Note that there is no retransmission at node $A_0$ and $\overline{r}_0=1$.

Our model represents a special case of interacting queues, which are notoriously difficult to analyze~\cite{Ephremides09}. To make the analysis
tractable, we \emph{assume} that:
\begin{enumerate}
\item Packet transmissions and retransmissions at each uncongested node $A_i$ form a Poisson process with rate~$\overline{r}_i
    \lambda$.\label{poisson arrival}
\item The probability that a packet transmitted by node $A_i$ collides is independent of previous attempts. This probability is denoted
    $p_i$.\label{independence}
\end{enumerate}

Though the assumption of Poisson retransmissions is not fully consistent with the Wi-Fi protocol, it is similar to the ``random-look'' model used by Kleinrock and Tobagi in their analysis of (single hop) random access
networks~\cite{kleinrock1975packet} (see also~\cite{bertsekas1992data}[Ch.\ 4]).
The simulations do not incorporate the simplifications used to make the analysis tractable, yet lead to the same effects.
 We stress that beside these assumptions, the rest of our analysis is exact.

\subsection{Iterative analysis of the utilization}
\label{model}

Our goal is to find the utilization at each node $i \geq 0$ and in the limit as $i \to \infty$. We consider the same scenario as in our simulations,
whereby node $A_0$ (the attacker) varies its traffic load
\begin{equation}
\rho_0 \triangleq \lambda_0 T,
\end{equation}
while all other nodes $A_i$ ($i \geq 1$) have the same traffic load
\begin{equation} \label{definition of rho}
\rho \triangleq \lambda_i T,
\end{equation}
where $0 < \rho < 1$.  We aim to understand if and how changes in the value of $\rho_0$ affect the utilization of nodes that are located far away as
function of the parameters $\rho$ and $R$.

First, we get the utilization at node $A_0$:
\begin{equation}
u_0 = \min\{\rho_0, 1\}.
\end{equation}

We next develop an iterative procedure to derive $u_{i+1}$ from $u_i$.
From (\ref{definition of utilization}) and (\ref{definition of rho}),
\begin{equation} \label{definition of utilization with rho}
u_{i+1}=\min\{ \overline{r}_{i+1} \rho, 1\}.
\end{equation}

We first relate $\overline{r}_{i+1}$ to $p_{i+1}$, the probability that a packet transmitted by node $A_{i+1}$ collides.
Based on Assumption~\ref{independence}, the probability that a packet is successfully received after $1 \leq r \leq R$ attempts is
 $(1 - p_{i+1}) (p_{i+1})^{r-1}$ while the probability that a packet  fails to be received after $R$ attempts is $(p_{i+1})^R $. 
 Hence, the
mean retry count at node $A_{i+1}$ is
\begin{eqnarray} \label{LP:average retry count without retry limit}
\overline{r}_{i+1} & = &  \sum_{r=1}^{R} r  \cdot (1 - p_{i+1}) \cdot  (p_{i+1})^{r-1} +R  \cdot (p_{i+1})^R \nonumber \\
& = & \sum_{r=1}^{R} (p_{i+1})^{r-1}.
\end{eqnarray}

We next relate  $p_{i+1}$ to $u_i$. First, suppose $u_i < 1$ (i.e., node $A_i$ is uncongested).   
 Assume that node $A_{i+1}$ starts a packet transmission (or retransmission) at some arbitrary time $t=t'$. We compute $p_{i+1}$ by
conditioning on whether or not node $A_i$ is transmitting at time $t'$.  Note that due the Poisson Arrivals See Time Averages (PASTA) property, the
transmission state of node $A_i$ at time $t=t'$ is the same as at any random point of time. 

If node $A_i$ transmits at time $t'$, which occurs with probability $u_i$, then the packet transmitted by node $A_{i+1}$ collides with probability 1. If node $A_i$
does not transmit at time $t'$, which occurs with probability $1-u_i$, then a collision occurs only if  node $A_i$ starts a transmission during the interval
$[t',t'+T]$. Since the packet inter-arrival time on the channel is exponentially distributed with mean $\overline{r}_i T$, such an event
occurs with probability
\begin{equation}
(1 - \mathrm{e}^{-\overline{r}_i \lambda_i T}) = (1 - \mathrm{e}^{-u_i}),
\end{equation}
based on Assumption~\ref{poisson arrival}. Therefore, the unconditional probability that a packet transmitted by node
$A_{i+1}$ collides is
\begin{eqnarray} \label{LP: f(u)}
  p_{i+1} &  = & 1 \cdot u_i + (1 - \mathrm{e}^{-u_i})\cdot(1-u_i) \nonumber \\
  & = & 1 - \mathrm{e}^{-u_i}(1-u_i).
\end{eqnarray}

Next, suppose $u_i = 1$ (i.e., node $A_i$ is congested). In that case, all the transmissions by node $A_{i+1}$ collide and $p_{i+1}=1$. We note that
(\ref{LP: f(u)}) still provides the correct result.

Putting (\ref{definition of utilization with rho}), (\ref{LP:average retry count without retry limit}), and
(\ref{LP: f(u)}) together, we obtain
\begin{equation}  \label{LP: convergence of the phase}
u_{i+1}  =  \min \left\lbrace \rho \sum_{r=1}^{R} \left( 1 - \mathrm{e}^{-u_{i}}(1-u_{i}) \right) ^{r-1}, 1 \right\rbrace.
\end{equation}

\subsection{Limiting behaviour of the utilization}
We next analyze the limiting behaviour of the iteration given by (\ref{LP: convergence of the phase}). The sequence $(u_i)_{i=0}^{\infty}$
corresponds to a discrete non-linear dynamical system~\cite{lynch2004dynamical}. Such systems are generally complex as they may converge to a point,
to a cycle  (i.e., they exhibit periodic behaviour), or not converge at all (i.e., they exhibit chaotic behaviour).

The main result of this section is to show that the sequence $(u_i)_{i=0}^{\infty}$ always converges to a point. However, the limit depends on the
initial utilization $u_0$.

To simplify notation, we define the function
\begin{equation}
f(u_{i}) \triangleq \rho \sum_{r=1}^{R} \left( 1 - \mathrm{e}^{-u_{i}}(1-u_{i}) \right) ^{r-1}.
\end{equation}

We then rewrite (\ref{LP: convergence of the phase}) as follows:
\begin{equation} \label{LP: simplification of convergence of the phase}
u_{i+1}  =  \min \left\lbrace f(u_{i}), 1 \right\rbrace.
\end{equation}

We say that $\omega \in [0,1]$ is a \emph{fixed point} of (\ref{LP: simplification of convergence of the phase})
if
\begin{equation} \label{number of fixed points plus}
\omega  =  \min \left\lbrace f(\omega), 1 \right\rbrace.
\end{equation}

Suppose (\ref{number of fixed points plus}) has $K$ different fixed points (Theorem~\ref{Thm: three stages} in the sequel will show that $K \geq
1$). We denote by $\Omega$ the ordered set of all the fixed points of (\ref{number of fixed points plus}). That is,
\begin{equation}
\Omega \triangleq\{\omega_1,\ldots,\omega_k,\ldots, \omega_K\},
\end{equation}
where $\omega_1<\ldots<\omega_k<\ldots< \omega_K$.

We are next going to show that for any $u_0\in [0, 1]$, the limit of the sequence $(u_i)_{i=0}^{\infty}$ is one of the elements in $\Omega$. To
prove this result, we will use the following lemma.

\begin{lemma} \label{lemma: f(w)>w}
Let $u, u' \in (\omega_k, \omega_{k+1})$, where $k \in \{1,\ldots,K-1\}$.
 If $f(u)>u$, then $f(u') > u'$. If $f(u)<u$, then $f(u') < u'$.
\end{lemma}
\begin{IEEEproof}
The proof goes by contradiction. Let $u, u' \in (\omega_k, \omega_{k+1})$. Suppose $f(u)>u$ and $f(u')<u'$. Since $f$ is continuous in $(\omega_k,
\omega_{k+1})$, then by the intermediate-value theorem there exists a point $u''$ between $u$ and $u'$ such that $f(u'') = u''$ . Thus, $u''$ is a
fixed point of (\ref{number of fixed points plus}). This contradicts the fact that no fixed point exists between $\omega_k$ and $\omega_{k+1}$.
\end{IEEEproof}


We now present the main result of this section.
\begin{theorem} \label{thm: convergence of the fixed point}
\leavevmode
\begin{enumerate}
\item Let $u_0 \in (\omega_k, \omega_{k+1})$, where $k \in \{1,\ldots,K-1\}$.
If $f(u_{0}) > u_{0}$, the sequence $(u_i)_{i=0}^{\infty}$ converges to $\omega_{k+1}$.
If $f(u_{0}) < u_{0}$, the sequence $(u_i)_{i=0}^{\infty}$ converges to $\omega_{k}$.
\item If $u_0 \in [0, \omega_1)$,  the sequence $(u_i)_{i=0}^{\infty}$ converges to $\omega_1$.
\item If $\omega_K<1$ and $u_0 \in (\omega_K, 1]$,  the sequence $(u_i)_{i=0}^{\infty}$ converges to $\omega_K$.

\end{enumerate}
\end{theorem}
\begin{IEEEproof}
\leavevmode
\begin{enumerate}
\item Let $\omega_k < u_0 < \omega_{k+1}$, where $k \in \{1,\ldots,K-1\}$. Since
$p_i \in (0, 1)$. Therefore,
the function $f$ is continuous and monotonically increasing,
$f(\omega_k) < f(u_0) < f(\omega_{k+1})$. Hence, according to (\ref{LP: simplification of convergence of the phase}) and (\ref{number of fixed
points plus}), we get
\begin{equation} \label{w_k < u_1 < w_k+1}
\omega_k \leq u_1 \leq \omega_{k+1}.
\end{equation}

Now, suppose \mbox{$u_1= f(u_{0}) > u_{0}$}.
If $u_1 = \omega_{k+1}$, then the result is proven.
If $u_1 < \omega_{k+1}$, then by Lemma~\ref{lemma: f(w)>w} and  Equation~(\ref{w_k < u_1 < w_k+1}), we have $u_2 = f(u_1) > u_1$. Applying the
same argument inductively, either there exists some value $M \geq 2$ such that $u_i=\omega_{k+1}$ for all $ i \geq M$, or the sequence
$(u_i)_{i=0}^{\infty}$ is monotonically increasing and upper bounded by $\omega_{k+1}$. According to the monotone convergence theorem, the
sequence converges. Since there is no other fixed point between $u_0$ and $\omega_{k+1}$ and $f$ is continuous, the sequence
$(u_i)_{i=0}^{\infty}$ must converge to $\omega_{k+1}$. The case $u_1 = f(u_{0}) < u_{0}$ is handled similarly.

\item Similar to Lemma~\ref{lemma: f(w)>w}, one can show that if there exists $u \in [0, \omega_1)$ such that  $f(u) >u$, then $f(u')>u'$ for all
    $u' \in [0, \omega_1)$.  Since $f(0)=\rho>0$, the sequence $(u_i)_{i=0}^{\infty}$ converges to $\omega_1$.

\item This is handled similarly to case 2.
\end{enumerate}
\end{IEEEproof}


\subsection{Phase transition analysis}
\label{The phase transition phenomenon}
In the previous section, we showed that the limit of the sequence of node utilizations $(u_i)_{i=0}^{\infty}$ must be one of the fixed points in the
set $\Omega$.  A phase transition represents a situation where a small change of $u_0$ leads to an abrupt  change of the limit. Specifically, we
focus on the case when the limit jumps to 1. Formally:
\begin{definition}[Network congestion] A network is said to be \textit{congested} if $(u_i)_{i=0}^{\infty}$ converges to $1$. Else, the network is
said to be \textit{uncongested}.
\end{definition}
\begin{definition}
[Phase transition] A network experiences a phase transition if there exists a fixed point $\omega \in \Omega$, such that if
$u_0 < \omega$ the network is uncongested, and if $u_0 > \omega$ the network is congested. We refer to $\omega$ as the phase transition point.
\end{definition}
We note that a phase transition can possibly occur only if $\omega_K=1$, since otherwise the network is never  congested, irrespective of $u_0$.


A network must fall in one of the following  three regimes:
\begin{enumerate}
\item The network is uncongested for all $u_0 \in [0,1]$.
\item The network is congested for all $u_0 \in [0,1]$.
\item A phase transition occurs.
\end{enumerate}
Our goal in the following is to determine what regime prevails under different network parameters.

%
%
%
%
%

For this purpose, we investigate the existence and properties of solutions of (\ref{number of fixed points plus}).
First, we investigate the case $\omega = 1$.
%
\begin{lemma} \label{lemma:overload fixed point is stable}
If $\rho > 1/R$, then
\begin{enumerate}
\item $\omega_K=1$.
\item If $K =1 $, then for all $u_0 \in [0,\omega_K]$ the sequence $(u_i)_{i=0}^{\infty}$ converges to $\omega_K$.
\item If $K \geq 2$, then for all $u_0 \in (\omega_{K-1},\omega_K]$ the sequence $(u_i)_{i=0}^{\infty}$ converges to $\omega_K$.
\end{enumerate}
\end{lemma}
\begin{IEEEproof}
\leavevmode
\begin{enumerate}
\item Let $\rho \geq 1/R$. We compute the RHS of (\ref{number of fixed points
plus}) at $\omega=1$ and obtain  $\min\{ f(1), 1\} = \min\{ R\rho, 1\} =1$, which proves that a fixed point indeed exists at $\omega=1$.

\item If $\rho > 1/R$, then $f(1)=R \rho>1$. Since $f(1)>1$, then for all $u_0 \in (0, \omega_K)$ , we have $f(u_0) > u_0$, based on an argument
    similar to Lemma~\ref{lemma: f(w)>w}, and the sequence $(u_i)_{i=0}^{\infty}$ converges to $1$, following an argument similar to
    Theorem~\ref{thm: convergence of the fixed point}.

\item This is handled similarly to Part 2.
\end{enumerate}
\end{IEEEproof}

Lemma~\ref{lemma:overload fixed point is stable} indicates that the sequence $(u_i)_{i=0}^{\infty}$ can converge to 1 (depending on $u_0$), if $
\rho > 1/R$. Besides this special case, (\ref{number of fixed points plus}) can be rewritten
\begin{equation} \label{LP: the fixed point}
f(\omega) = \omega.
\end{equation}
We look for solutions of (\ref{LP: the fixed point}) that belong to the interval $[0,1]$. Each such solution is an element of $\Omega$.

Equation (\ref{LP: the fixed point}) is difficult to work with because it contains two unknown variables, $\rho$ and $R$. To circumvent this
difficulty, we introduce the function
\begin{equation} \label{lambdaT}
h_R(\omega) \triangleq \frac{\rho \omega}{f(\omega)} = \frac{\omega}{\sum_{r=1}^{R}\left(1 - \mathrm{e}^{-\omega}(1-\omega)\right)^{r-1}}.
\end{equation}
For each value of $\rho$, the solutions of (\ref{LP: the fixed point}) must satisfy
\begin{equation} \label{lambdaT=rho}
h_R(\omega) = \rho.
\end{equation}
We denote the maximum of $h_R(\omega)$
by
 \[h^{max}_R \triangleq \max_{0 \leq \omega \leq 1}h_R(\omega).\]

The following theorem establishes the prevailing network regimes for different parameters.
\begin{theorem} \label{Thm: three stages}
\leavevmode
\begin{enumerate}
\item If $\rho < 1/R$, then the network is uncongested for all $u_0 \in [0,1]$.
\item If $h^{max}_R > 1/R$ and $1/R < \rho < h^{max}_R$, then a phase transition occurs and the phase transition point is  $\omega_{K-1}$.
\item If $\rho > h^{max}_R$, then the network is congested for all $u_0 \in [0,1]$.

\end{enumerate}
\end{theorem}
\begin{IEEEproof}
\leavevmode
\begin{enumerate}
\item If $\rho < 1/R$, then $R \rho < 1$ and the utilization of each node is always less than 1. Hence, for any $u_0 \in [0,1]$,  the network is
    always uncongested. Note that since $h_R(0)=0$, $h_R(1)=1/R$, and $h_R$ is continuous,  (\ref{lambdaT=rho}) must have at least one solution
    (i.e., at least one fixed point exists).

\item Let $\rho \in (1/R, h^{max}_R)$. We know that $h_R(0) = 0$ and $h_R(1)=1/R$. Since the function $h_R$ is continuous,  (\ref{lambdaT=rho})
    must have at least one solution (i.e, at least one fixed point strictly smaller than 1 exists). Also,  because $\rho > 1/R$, a fixed point
    point at $\omega=1$ exists (i.e., $\omega_K=1$), by Part 1 of Lemma~\ref{lemma:overload fixed point is stable}.  Thus, there are $K \geq 2$
    fixed points.

By Part 3 of Lemma~\ref{lemma:overload fixed point is stable}, the sequence $(u_i)_{i=0}^{\infty}$ converges to $\omega_K$ for all $u_0 \in
(\omega_{K-1},\omega_K]$.  Moreover, by Theorem~\ref{thm: convergence of the fixed point}, the limit of the sequence $(u_i)_{i=0}^{\infty}$ is no
larger than $\omega_{K-1}$ for all $u_0 \leq \omega_{K-1}$. Hence, a phase transition exists at $\omega_{K-1}$.

\item If $\rho > h^{max}_R$, then (\ref{LP: the fixed point}) has no solution.  Moreover, since $\rho > h^{max}_R \geq h_R(1) = 1/R$, we get $\rho
    > 1/R$. By Parts 1 and 2 of Lemma~\ref{lemma:overload fixed point is stable}, the sequence $(u_i)_{i=0}^{\infty}$ converges to 1 for any $u_0
    \in [0,1]$, and the network is always congested.

\end{enumerate}
\end{IEEEproof}

\begin{figure}[!t]
\centering
\subfloat[$R=4$]
{\includegraphics[width=1.15in]{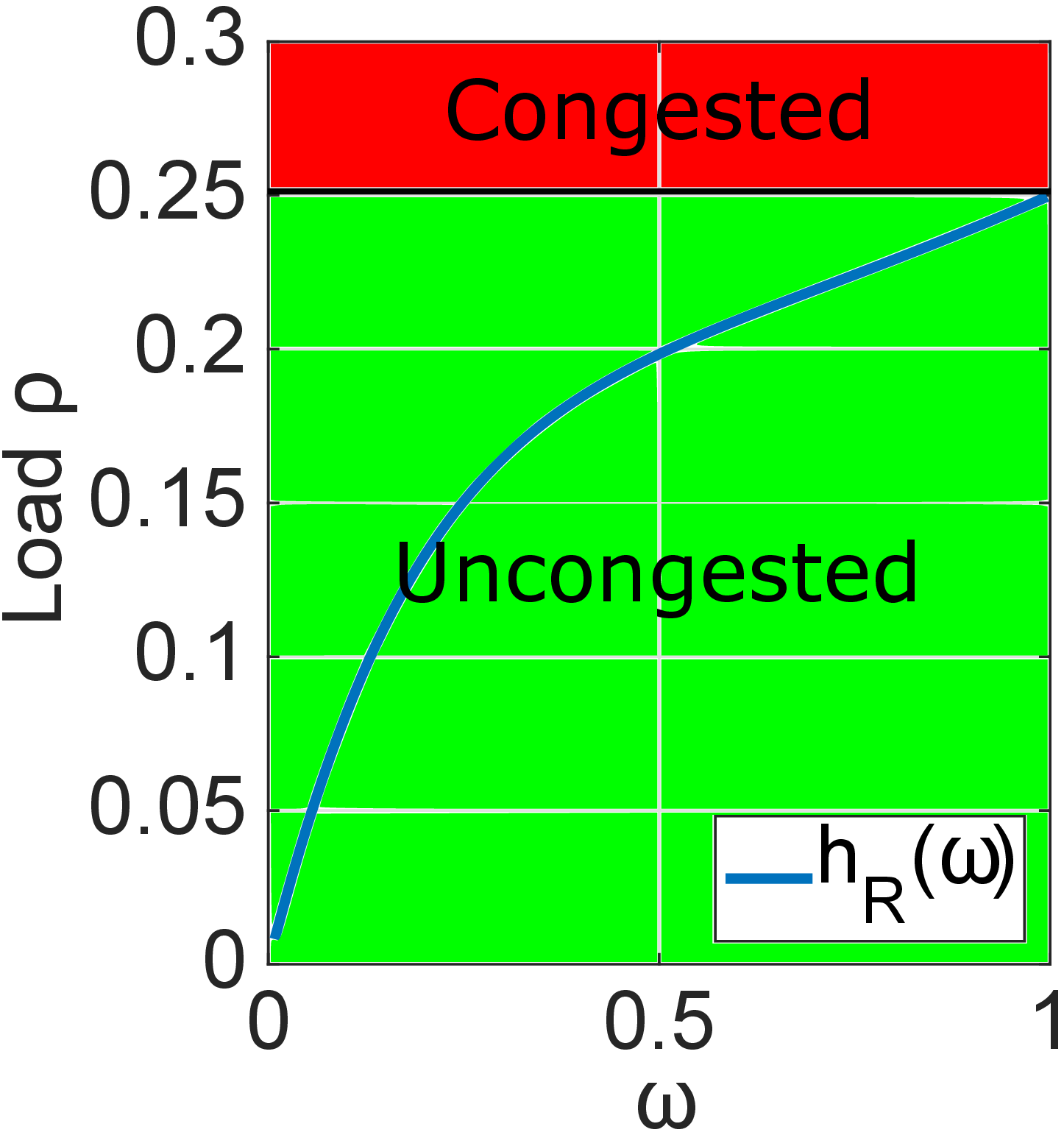}
\label{R is 4}}
\subfloat[$R=7$]
{\includegraphics[width=1.15in]{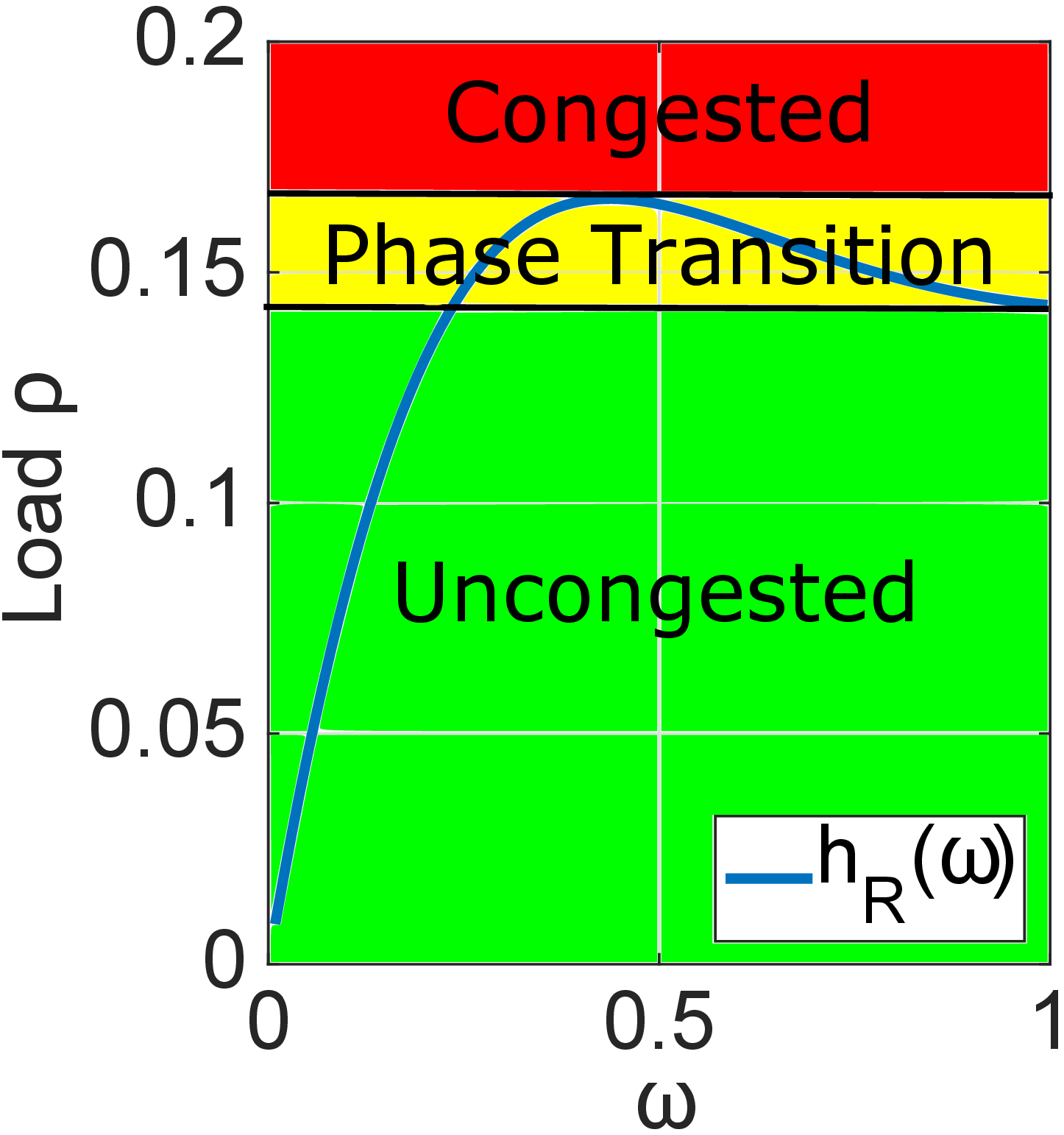}
\label{R is 7}}
\subfloat[$R=10$]
{\includegraphics[width=1.15in]{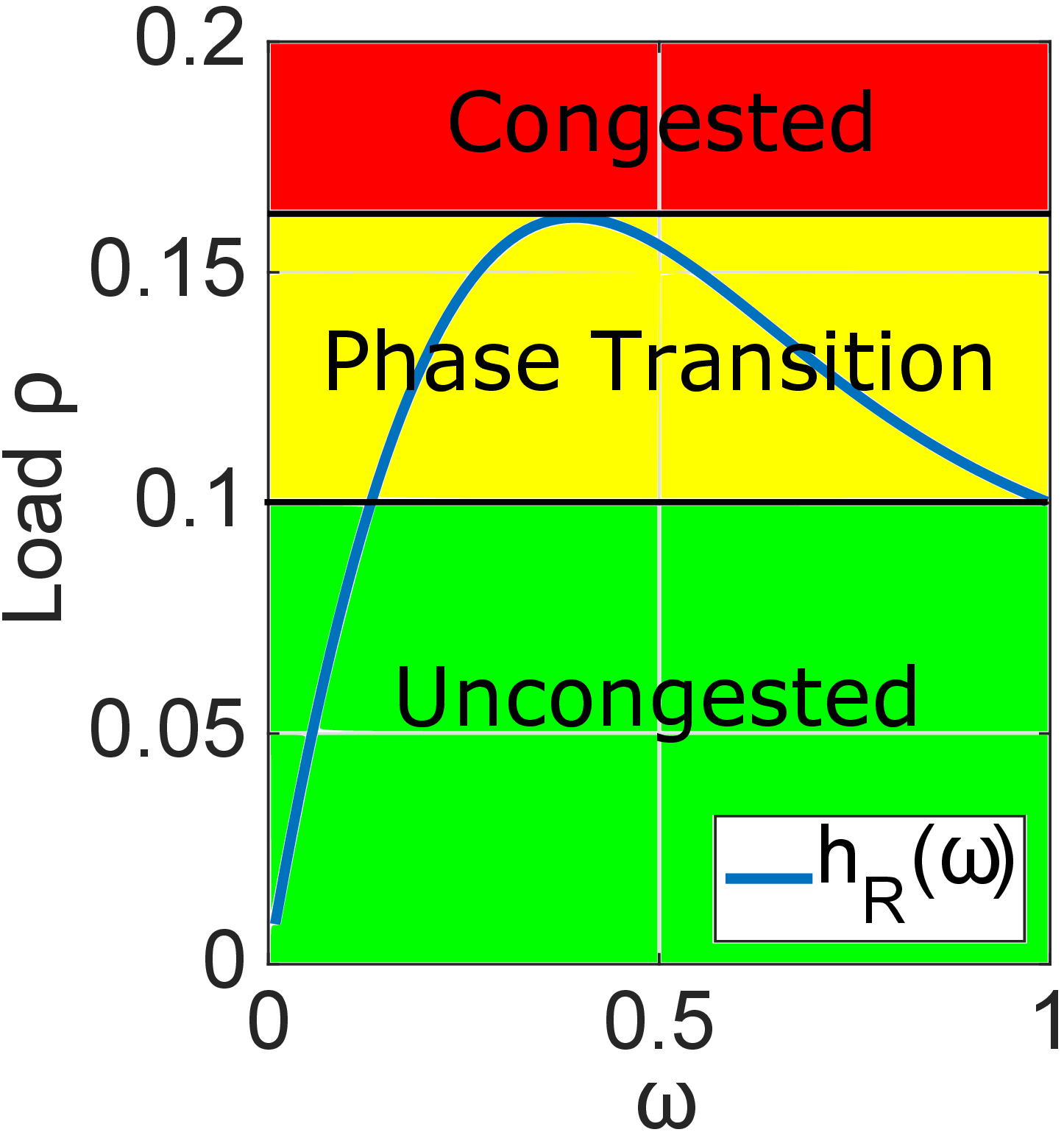}
\label{R is 10}}
\caption{Illustration of the different network regimes for different values of $R$. For each value of~$\rho$, the fixed points are the solutions of
$h_R(\omega)= \rho$. In addition, the fixed point $\omega =1$ always exists when $\rho > 1/R$. A phase transition region exists if the maximum of
$h_R(\omega)$, $h_R^{max}$, is strictly greater than $h_R(1) = 1/R$.}
\label{number_of_fixed_points}
\end{figure}

We next illustrate Theorem~\ref{Thm: three stages} for different values of $R$, using  Figure~\ref{number_of_fixed_points}.
First, consider $R = 4$ as shown in Figure~\ref{number_of_fixed_points}(a). Since $h^{max}_R = 1/R=0.25$, there exists no traffic load $\rho$ for
which a phase transition exists. Either the network is always uncongested (for $\rho < 1/R$), or it is always congested (for $\rho > 1/R$).

Next, consider $R = 7$ as shown in Figure~\ref{number_of_fixed_points}(b). There, $h^{max}_R = 0.166 > 1/R = 0.143$. Hence, a phase transition
occurs if $\rho \in (0.143, 0.166)$. For instance, consider the case $\rho = 0.15$. Then, the equation $h_R(\omega)=\rho$ has two solutions.
Including the fixed point $\omega=1$ (since $\rho > 1/R)$, the set $\Omega$ has $K=3$  fixed points: $\{ \omega_1=0.265, \omega_2=0.777,
\omega_3=1\}$. Hence, by Theorem~\ref{Thm: three stages}, the network is uncongested if $u_0 < 0.777$, and congested if $u_0 > 0.777$.

The case $R=10$ also has a phase transition region, as shown in Figure~\ref{number_of_fixed_points}(c). Furthermore, the size of this region is
larger since $(1/R, h^{max}_R) = (0.1, 0.162)$.


\subsection{Sufficient condition for phase transition}
\label{The sufficient condition for phase transition}
In the previous section, we showed that a phase transition exists in the region $1/R < \rho < h^{max}_R$, if $h^{max}_R > 1/R$. In this section, we
derive an explicit lower bound on $h^{max}_R$, which provides a simple  condition for the existence of a phase transition. First, we establish a
relationship between the derivatives of $h_{R}(\omega)$ for different values of $R$, but a given value of $\omega$.

\begin{lemma} \label{unstable then unstable}
For $\omega \in [0,1]$, if there exists $R^* \geq 1$ such that $h_{R^*}'(\omega) \leq 0$, then $h_R'(\omega) \leq 0$ for all $R>R^*$.
\end{lemma}
\begin{IEEEproof}
Let $\omega \in [0,1]$.
Since
\begin{equation}\label{h^-1}
\left(h_R^{-1}(\omega)\right)' = - \frac{h_R'(\omega)}{h_R(\omega)^2},
\end{equation}
the sign of $h_R'(\omega)$ is opposite to $\left(h_R^{-1}(\omega)\right)'$. Hence, we investigate the sign of
\begin{equation} \label{1/h'(w)}
\left(h_R^{-1}(\omega)\right)' = \sum_{r=1}^{R}\Psi_r'(\omega),
\end{equation}
where
\begin{equation}\label{Psi_r}
\Psi_r(\omega) \triangleq \frac{\left(1 - \mathrm{e}^{-\omega}(1-\omega)\right)^{r-1}}{\omega}.
\end{equation}

We check the sign of each term $\Psi_r'(\omega)$ in~(\ref{1/h'(w)}), for $r \in \{1, 2, \ldots, R\}$.
For $r=1$, we have
\[\Psi_1'(\omega)=\frac{d}{d \omega}\left(\frac{1}{\omega}\right)=-\frac{1}{\omega^2}<0.\]

For $r\geq2$, we have
\begin{equation}\label{Psi_r'} \Psi_r'(\omega) =-\frac{e^{-\omega} \left(1 - \mathrm{e}^{-\omega}(1-\omega)\right)^{r-2}\Phi_r(\omega)}{\omega^2},
\end{equation}
where \[\Phi_r(\omega) \triangleq -1+e^\omega+(3-2r)\omega+(r-1)\omega^2.\] Clearly, the terms $e^{-\omega}$, $\left(1 -
\mathrm{e}^{-\omega}(1-\omega)\right)^{r-2}$ and $\omega^2$ in~(\ref{Psi_r'}) are all positive. Thus, the signs of $\Phi_r(\omega)$ and
$\Psi_r'(\omega)$ are opposite.

We next investigate the signs of the first and second derivatives of the function $\Phi(\omega)$. We have
\begin{eqnarray}
\Phi_r'(\omega) &  = &  e^\omega+3-2r+2(r-1)\omega, \label{Phi_r'}\\
\Phi_r''(\omega)&  =  & e^\omega+2(r-1)>0, \label{Phi_r''}
\end{eqnarray}
for all $\omega \in [0,1]$ and $r \geq 2$.
From (\ref{Phi_r''}), we find that $\Phi_r'(\omega)$ is monotonically increasing with $\omega$.

For any $r\geq2$, we obtain from (\ref{Phi_r'}) that
\begin{eqnarray}
\Phi_r'(0) & = & 4-2r, \\
\Phi_r'(1) & = & e+1.
\end{eqnarray}

We distinguish between three possible cases regarding the sign of $\Phi_r(\omega)$:
\begin{enumerate}
	\item For $r=2$, $\Phi_2'(0)=0$. Hence, $\Phi_2'(\omega)>0$. The function $\Phi_2(\omega)$ is monotonically increasing with $\omega$. Since
$\Phi_2(0)=e-1>0$, $\Phi_2(\omega)$ is always positive.
	\item For $r=3$, $\Phi_3'(0)<0$. The function $\Phi_3(\omega)$ first decreases then increases as $\omega$ increases from 0 to 1. Since
$\Phi_3(0)=0$ and $\Phi_3(1)>0$, the sign of the function $\Phi_3(\omega)$ turns from negative to positive as $\omega$ increases from $0$ to $1$.
	\item For $r>3$, $\Phi_r'(0)<0$. The function $\Phi_r(\omega)$ first decreases then increases as $\omega$ increases from 0 to 1. Since
$\Phi_r(0)=0$ and $\Phi_r(1)<0$, the sign of the function $\Phi_r(\omega)$ is always negative.
\end{enumerate}

Therefore, by~(\ref{1/h'(w)}), for any given $\omega \in [0,1]$, the sign of the function $\Phi_r(\omega)$ turns from being positive to being
negative as $r$ increases. Equivalently, the sign of the function $\Psi_r'(\omega)$ turns from being negative to being positive as $r$ increases.

Thus, by~(\ref{1/h'(w)}), if $\left(h_{R}^{-1}(\omega)\right)'$ is nonnegative for $R=R^*$, then it is also nonnegative for all $R \geq R^*$.
Equivalently, by~(\ref{h^-1}), if $\left(h_R^{-1}(\omega)\right)'$ is nonpositive for $R=R^*$, then it is also nonpositive for all $R \geq R^*$,
which completes the proof.
\end{IEEEproof}

Consider the function $h_R(\omega)$ as $R\rightarrow\infty$:
\begin{eqnarray} \label{lim R infty h_R}
h_\infty(\omega) & = & (1-\left(1 - \mathrm{e}^{-\omega}(1-\omega)\right))\omega \nonumber\\
& = & e^{-\omega}(1-\omega)\omega,
\end{eqnarray}
and its derivative
\begin{equation} \label{lim R infty h_R'}
h_\infty'(\omega) =  e^{-\omega}(1-3\omega+\omega^2).
\end{equation}
The next corollary is the logical transposition of Lemma \ref{unstable then unstable}.
\begin{corollary} \label{stability when R infty}
If $h_{\infty}'(\omega) \geq 0$, then $h_R'(\omega) \geq 0$ for all $R \geq 1$.
\end{corollary}

The following lemma establishes that the function $h_R(\omega)$ is always strictly increasing in the interval $[0, \overline{\omega})$, where
\begin{equation}\label{overline_omega}
\overline{\omega} \triangleq \frac{3-\sqrt{5}}{2}.
\end{equation}
\begin{lemma}
\label{stable fixed point}
Let $0 \leq \omega<\overline{\omega}$. Then, $h_R'(\omega)>0$, for all $R \geq 1$.
\end{lemma}
\begin{IEEEproof}
Let the function $h_\infty(\omega)$ and its derivative $h_\infty'(\omega)$ be defined as in (\ref{lim R infty h_R}) and (\ref{lim R infty h_R'}),
respectively.
Since $e^{-\omega}$ is always positive, $h_\infty'(\omega)$ has the same sign as $(1-3\omega+\omega^2)$.
The unique root of $(1-3\omega+\omega^2)=0$ for $\omega \in [0,1]$ is $\bar{w}$ as defined in (\ref{overline_omega}).

Thus, $(1-3\omega+\omega^2)$ is positive when $0 \leq \omega<\overline{\omega}$,
and so is $h_\infty'(\omega)$. By Corollary \ref{stability when R infty}, $h_R'(\omega)>0$ for $0 \leq \omega<\overline{\omega}$ and for all $R \geq
1$.

\end{IEEEproof}

The consequence of Lemma~\ref{stable fixed point} is that for all $R \geq 1$,
\begin{equation} \label{h_max_h_omega}
h^{max}_R \geq h_R(\overline{\omega}).
\end{equation}
This equation provide a lower bound on $h^{max}_R$ that can easily be computed. We then obtain the following sufficient condition for the existence
of phase transition.

\begin{theorem}
\label{Thm:range of lambda T}
Let $\overline{\omega}$ be defined as in (\ref{overline_omega}) and suppose $h_R(\overline{\omega})> 1/R$. Then, a phase transition is guaranteed to
exist for any $\rho \in (1/R, h_R(\overline{\omega}))$.
\end{theorem}
\begin{IEEEproof}
From Theorem \ref{Thm: three stages}, we know that a phase transition exists if $1/R < \rho< h^{max}_R$. By~(\ref{h_max_h_omega}) and the assumption
that $h_R(\overline{\omega})> 1/R$, the proof follows.
\end{IEEEproof}


The next theorem establishes an even more explicit lower bound on $h^{max}_R$.


\begin{theorem}
\label{Thm:specific range of lambda T}
Let $h_\infty(\omega)$ and $\overline{\omega}$ be defined as in (\ref{lim R infty h_R}) and (\ref{overline_omega}), respectively. Then,
$h^{max}_R \geq h_\infty(\overline{\omega}) \simeq 0.161$.
%
\end{theorem}
\begin{IEEEproof}
By~(\ref{lambdaT}),
\begin{eqnarray}
h_R(\overline{\omega}) & = & \frac{\omega}{\sum_{r=1}^{R}(1-e^{-\omega}(1-\omega))^{r-1}} \nonumber\\
& > & \frac{\omega}{\sum_{r=1}^{\infty}(1-e^{-\omega}(1-\omega))^{r-1}} = h_\infty(\overline{\omega}). \label{hinf_ineq}
\end{eqnarray}
Thus, by~(\ref{h_max_h_omega}) and ~(\ref{hinf_ineq}), $h^{max}_R > h_\infty(\overline{\omega}) \simeq 0.161$. Note that this bound is
asymptotically tight as $R \to \infty$ since $h^{max}_{\infty} = h_\infty(\overline{\omega})$.
\end{IEEEproof}

From Theorems~\ref{Thm: three stages} and~\ref{Thm:specific range of lambda T}, it follows that a phase transition exists if $1/R < 0.161$. Hence:
\begin{corollary}\label{R>=7}
A phase transition is guaranteed to exist for $R\geq7$ and $\rho \in [1/R,0.161]$.
\end{corollary}
We note that the lower bound on $h^{max}_R$ is quite tight. For instance, $h^{max}_7=0.166$. Moreover, $h^{max}_R$ decreases with $R$ (this follows
from~(\ref{lambdaT}), since for any $\omega \in [0,1]$ the denominator increases as $R$ gets larger).

\subsection{Stability of fixed points}
\label{Stability of Fixed point}
In this subsection, we use stability theory to shed further light into the limiting behaviour of the sequence $(u_i)_{i=0}^{\infty}$.
Specifically, the sequence $(u_i)_{i=0}^{\infty}$ converges to \emph{stable} fixed points of $\Omega$ and diverges from \emph{unstable} fixed points
of $\Omega$. We will show that the stability of the fixed points of~(\ref{LP: the fixed point}) are determined by the sign of $h_R'(\omega)$ at
those points.

Informally, a fixed point $\omega$ is stable (or an  \emph{attractor}),  if there exists a domain containing $\omega$, such that if $u_0$ belongs to
that domain, then $(u_i)_{i=0}^{\infty}$ converges to $\omega$.
\begin{definition}[Stability of a fixed point] \label{def: stable fixed point}
Let $u_0 \in [0,1]$. A fixed point $\omega \in \Omega$ is \textit{stable} if there exists $\epsilon>0$ such that if $|u_0-\omega|< \epsilon$, the
sequence $(u_i)_{i=0}^{\infty}$ converges to $\omega$.
It is \textit{unstable} if for all $u_0 \neq \omega$ the sequence $(u_i)_{i=0}^{\infty}$ does not converge to $\omega$.
\end{definition}

Recall that according to Lemma \ref{lemma:overload fixed point is stable}, a special fixed point of (\ref{number of fixed points plus}) exists at
$\omega = 1$, if $\rho > 1/R$. According to Definition~\ref{def: stable fixed point}, this fixed point is stable.
Besides this special case, the rest of the fixed points satisfy Equation~(\ref{LP: the fixed point}).
To establish the stability of those fixed points, we will employ the following proposition.
\begin{proposition}[{\cite{lynch2004dynamical}}]\label{prop: stable fixed point}
Suppose that a continuously differentiable function $f$ has a fixed point $\omega$. Then, $\omega$ is stable if \mbox{$|f'(\omega)| < 1$} and
unstable if $|f'(\omega)| > 1$.
\end{proposition}

The next theorem provides a criterion to establish the stability of a fixed point $\omega \in \Omega$ with respect to the function $h_R(\omega)$.
\begin{theorem} \label{stability}
Consider a fixed point $\omega \in \Omega$, where $\omega < 1$. Then $\omega$ is stable if $h_R'(\omega) > 0$ and unstable if $h_R'(\omega) < 0$.
\end{theorem}
\begin{IEEEproof}
Let $\omega \in \Omega$. The derivative of $h_R(\omega)$ with respect to $\omega$ is
\begin{equation} \label{Eq: partial of h}
h_R'(\omega) =  \frac{1}{\Gamma(\omega)} -  \frac{\omega}{(\Gamma(\omega))^2} \cdot \Gamma'(\omega)  >  0,
\end{equation}
where
\begin{equation} \label{Eq: Gamma omega}
\Gamma(\omega) \triangleq \sum_{r=1}^{R}\left(1 - \mathrm{e}^{-\omega}(1-\omega)\right)^{r-1} = \frac{f(\omega)}{\rho}.
\end{equation}
If one can show that (\ref{Eq: partial of h}) implies $|f'(\omega)|<1$, then according to Proposition \ref{prop: stable fixed point}, the fixed
point $\omega$ is stable. We multiply both sides of (\ref{Eq: partial of h}) by $(\Gamma(\omega))^2$ and obtain
\begin{equation} \label{Eq: transform of partial of h}
\Gamma(\omega) - \omega \Gamma'(\omega) > 0.
\end{equation}
Using (\ref{Eq: Gamma omega}) and (\ref{LP: the fixed point}), we can rearrange (\ref{Eq: transform of partial of h}) as follows:
\begin{equation} \label{Gamma < 1 rho}
\Gamma'(\omega) < \frac{\Gamma(\omega)}{\omega} = \frac{f(\omega)}{\rho \omega} = \frac{1}{\rho}.
\end{equation}
From (\ref{Eq: Gamma omega}) and (\ref{Gamma < 1 rho}), we get \[f'(\omega) = \rho \Gamma'(\omega) < 1.\]
Since $f(\omega)$ is monotonically increasing with $\omega$, for $\omega \in [0,1]$, we conclude
\[0 < f'(\omega) < 1.\]
Hence, by Proposition \ref{prop: stable fixed point}, $\omega$ is a stable fixed point.

Similarly, $h_R'(\omega) < 0$ implies $f'(\omega) > 1$, which means that $\omega$ is unstable.

\end{IEEEproof}

\begin{figure}[!t]
\centering
\includegraphics[width=3.2in]{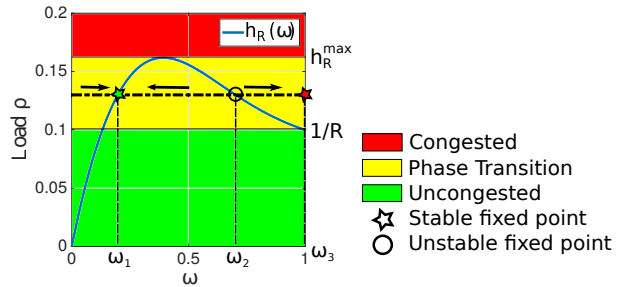}
\caption{Stability of fixed points with $R=10$. Given a load $\rho = 0.13$ (dash line), $\Omega$ contains three fixed points: $\omega_1 = 0.2$,
$\omega_2 = 0.7$ and $\omega_3=1$. The fixed point $\omega_1$ is stable because $h'_R(\omega_1) > 0$ and $\omega_2$ is unstable because
$h'_R(\omega_2) < 0$. The fixed point $\omega_3=1$ exists and is stable because $\rho > 1/R$. Therefore, the sequence $(u_i)_{i=0}^{\infty}$
converges to $\omega_1$ if $u_0 < \omega_2$, and to $\omega_3$ if $u_0 > \omega_2$.}
\label{stability example}
\end{figure}

We next show how the stability analysis of the fixed points helps to determine the limit of the sequence $(u_i)_{i=0}^{\infty}$. Consider, for
instance, the example shown in Figure \ref{stability example} with parameters $R=10$ and $\rho=0.13$. Under these parameters, $\Omega = \{\omega_1,
\omega_2, \omega_3\} = \{0.2, 0.7, 1\}$.

The fixed points $\omega_1$ and $\omega_2$ are the solutions of $h_R(\omega) = \rho$. According to Theorem \ref{stability}, $\omega_1$ is stable and
$\omega_2$ is unstable. The fixed point $\omega_3=1$ exists and is stable, since $\rho > 1/R$.

According to Theorem \ref{Thm: three stages}, $\omega_2$ is a phase transition point. Hence, the sequence $(u_i)_{i=0}^{\infty}$ converges to
$\omega_1$ if $u_0 < \omega_2$ and the network is uncongested. If $u_0 > \omega_2$, the sequence converges to $\omega_3$ and the network is
congested.

\subsection{Heterogeneous traffic load}
\label{sec:heter}
In previous subsections, we assumed that node $A_0$ varies its traffic load
$\rho_0$, but all other nodes $A_i$ ($i \geq 1$) have the same traffic load $\rho$.
We now relax this assumption and assume that nodes $A_i$ ($i \geq 1$) have different traffic loads
$\rho_i = \lambda_i T$.
We next prove that a phase transition still occurs, as long as all the traffic loads fall in the appropriate range.

\begin{theorem}
Suppose $h^{max}_R > 1/R$. If  $\rho_i \in (1/R, h^{max}_R)$ for all $i  \geq 1$, then a phase transition occurs.
\end{theorem}
\begin{IEEEproof}
Let $\rho_{max} = \max_{i \geq 1} \rho_i$ and $\rho_{min} = \min_{i \geq 1} \rho_i$.
According to Theorem~\ref{Thm: three stages}, the network is uncongested when $\rho_0=0$ and the load at each node $A_i$ is $\rho_{max} <
h_R^{max}$.
Hence, the network must remain uncongested when the load at each node $A_i$ is smaller than $\rho_{max}$.

Similarly, the network is congested when $\rho_0=1$ and the load at each node $A_i$ is $\rho_{min} > 1/R$.
Hence, it must remain congested when the load at each node $A_i$ is larger than $\rho_{min}$.
Thus, a phase transition occurs when $1/R < \rho_i < h_R^{max}$ for all $i  \geq  1$.
\end{IEEEproof}

\subsection{Comparison with simulation results}
\label{Mitigation}
We compare the results of our analysis with ns-3 simulations, for different settings of the retry limit $R$ and load $\rho$. For the simulations, we
consider an ad~hoc network composed of 41 pairs of nodes, as described in Section~\ref{EXP:Fixed bitrate}. 

\subsubsection{Region of phase transition}
\label{validation of Region of phase transition}
To check whether a phase transition exists for a given $R$, we run simulations both for $\rho_0 =0$ and $\rho_0 = 1$. If the node utilizations in
the limit (i.e., for node $A_{40}$) is the same in both cases, then we assume that there is no phase transition. If the limits are different, then a
phase transition exists.

\begin{figure}[!t]
\centering
\includegraphics[width=2.5in]{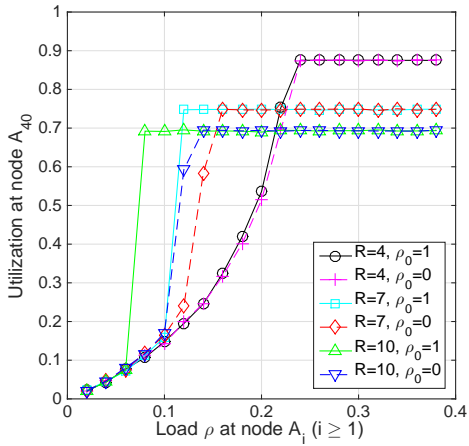}
\caption{Simulation of the limiting behaviour of the node utilization in a network of 41 pairs of nodes. For $R=4$, the limit is the same when
$\rho_0 = 0 $ and $\rho_0 = 1$, hence no phase transition is observed. However, for $R=7$ and $R=10$,
the limits are different, hence showing the existence of a region of load $\rho$ in which a phase transition occurs. }
\label{retry_limit}
\end{figure}

Figure~\ref{retry_limit} indicates that the existence of a phase transition is related to the retry limit, as predicted by our analysis. For the
case $R=4$, there is no phase transition, while a phase transition occurs in the cases $R=7$ and $R=10$. 
in our simulations for any $R \leq 6$.

The analysis also reasonably approximates the phase transition region. For $R=7$, the simulations show that a phase transition exists if $\rho \in
(0.12,0.16)$, while the analysis predicts $\rho \in (0.14, 0.17)$. For $R=10$, the simulation results are $\rho \in (0.08, 0.14)$ while the analysis
predicts $\rho \in (0.10, 0.16)$. We note that the size of the phase transition region increases with $R$, as predicted by the analysis.

\subsubsection{Heterogeneous traffic load}
\label{validation of Heterogeneous traffic load}
We next show the feasibility of a cascading DoS attack in
a network where the traffic load at different node is heterogeneous, in line with the analysis of Section~\ref{sec:heter}. 
Specifically,  the traffic load $\rho_i$ at each node $A_i$  ($i \geq 1$) is a continuous random variable that is uniformly distributed between 0.11 and
0.15.

\begin{figure}[!t]
\centering
\includegraphics[width=2.5in]{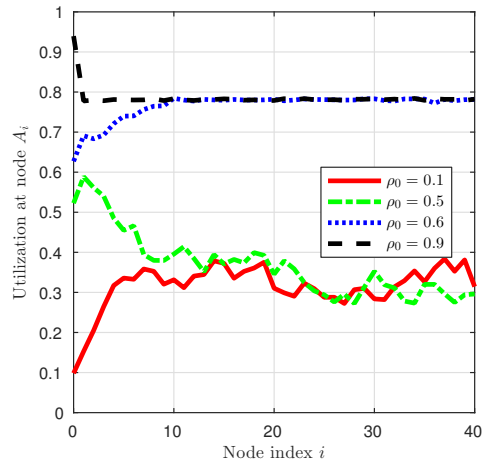}
\caption{Simulation with heterogeneous traffic load in a network with 41 pairs of nodes. The traffic load of nodes $A_i$ ($i \geq 1$) are uniformly
distributed between 0.11 and 0.15. For $R=7$, when the load $\rho_0$ changes from 0.5 to 0.6, the limiting behavior of the sequence of node utilizations differs, thus indicating the occurrence of phase transition.}
\label{CDoS-1Mbps-adhoc-UDP-randomrho-utilization}
\end{figure}

Figure~\ref{CDoS-1Mbps-adhoc-UDP-randomrho-utilization} shows the simulation results for retry limit $R=7$. 
When $\rho_0$, the load of node $A_0$,   is below 0.5, the network is uncongested and the utilizations of nodes $A_i$ oscillate around 0.35 as $i$ gets large. Note that the sequence does not converge to a fixed value due to the different traffic loads at the different nodes.
However, when $\rho_0$ exceeds 0.6, the sequence of node utilizations converges to its upper limit, implying that the network is congested. 

\section{Conclusion}
\label{Conclusion}
We describe a new type of DoS attacks against Wi-Fi networks, called cascading DoS attacks.
The attack exploits a coupling vulnerability due to hidden nodes.
The attack propagates beyond the starting location, lasts for long periods of
time, and forces the network to operate at its lowest bit rate. The attack can be started
remotely and without violating the IEEE 802.11 standard, making it
difficult to trace back.

We demonstrate the feasibility of such attacks, both through experiments on a testbed and extensive ns-3 simulations.
The simulations show that the attack is effective in networks operating under fixed and varying bit rates, as well
as ad hoc and infrastructure modes. We show that a small change
in the traffic load of the attacker can lead to a phase transition of the entire network, from uncongested state to congested state.

We develop an iterative analysis to characterize the sequence of node utilizations, and study its limiting behaviour. We show that the sequence
always converges to stable fixed points while an unstable fixed point represents a phase transition point. Based on the system parameters, we
identify when the system remains always uncongested, congested, or experiences a phase transition caused by a DoS cascading attack.

The analysis predicts that a phase transition occurs for $R \geq 7$ and provides a simple and explicit estimate of traffic load at each node under
which a phase transition occurs  (i.e., $\rho_i \in (1/R, 0.161)$ for all $i \geq  1$). The network is always congested when the traffic load is
above the phase transition regime and always uncongested when the traffic load is below the phase transition regime. Although the analysis is based
on some simplifying assumptions, the estimate is not far from the values observed in the simulations.

Exploiting the coupling vulnerability in different network configurations represents an interesting area for future work.
Experience in the security field indeed teaches that once a vulnerability is identified, more  potent attacks are subsequently discovered
(consider,
for instance, the history of attacks on WEP~\cite{tews2007breaking} and MD5~\cite{black2006study}). In our case, our simulations for ring topologies
indicate that the presence of a cycle in the topology could reinforce cascading DoS attacks, a result that warrants further investigations.

Several approaches are possible to mitigate cascading DoS attacks. First, one could enable the RTS/CTS exchange, although this solution has several
drawbacks, including major performance degradation under normal network operations, as mentioned in the Introduction. Devising a scheme that triggers RTS/CTS under certain circumstances (e.g., multiple consecutive packet losses) could be an interesting area for future research.
The second approach is to
lower the retry limit. However, this could also negatively impact performance. Other approaches include using short packets, collision-aware rate
adaptation algorithms, dynamic channel selection, and full-duplex radios. We leave the investigation and comparison of these
mitigation techniques as possible areas for future work.


\bibliographystyle{IEEEtran}
\bibliography{IEEEabrv,Cascading_DoS_Journal_v2.17}
%
%
%

\end{document}